\documentclass[aps,pra,twocolumn,a4paper,superscriptaddress,10pt]{revtex4-1}
\usepackage{graphicx}
\usepackage{amsmath}
\usepackage{amssymb}
\usepackage{mathrsfs}
\usepackage{amsfonts}
\usepackage{bm}
\usepackage{textgreek}
\usepackage{color}
\usepackage{xcolor}
\usepackage{paralist}
\usepackage{multirow}
\usepackage[inline]{enumitem}
\usepackage{url}

\usepackage[all]{xy}

\providecommand{\abs}[1]{\left\lvert#1\right\rvert}

\newcommand{\re}{\operatorname{Re}}
\newcommand{\im}{\operatorname{Im}}

\newcommand{\Heq}{{}^4\text{He}}

\newcommand{\br}{\mathbf{r}}

\newcommand{\vE}{\mathbf{E}}
\newcommand{\vB}{\mathbf{B}}
\newcommand{\vD}{\mathbf{D}}

\newcommand{\vH}{\mathbf{H}}

\newcommand{\bnabla}{\bm{\nabla}}

\newcommand{\di}{\text{d}}

\newcommand{\hD}{\hat{D}}

\newcommand{\hbr}{\mathbf{\hat{r}}}

\newcommand{\veo}{{\varepsilon_0}}
\newcommand{\ver}{{\varepsilon_r(\br)}}

\newcommand{\tla}{\tilde{\lambda}}

\newcommand{\tphi}{\tilde{\phi}}

\begin{document}

\title{Perturbation theory of optical resonances  of deformed  dielectric spheres}






\author{Andrea Aiello}
\email{andrea.aiello@mpl.mpg.de} 
\affiliation{Institute for Theoretical Physics, Department of Physics, University of Erlangen-N\"{u}rnberg, Staudtstrasse 7, 91058 Erlangen, Germany}
\affiliation{Max Planck Institute for the Science of Light, Staudtstrasse 2, 91058
Erlangen, Germany}

\author{Jack G. E. Harris}
\affiliation{Department of Physics, Yale University, New Haven, CT, 06520, USA}

\author{Florian Marquardt}
\affiliation{Institute for Theoretical Physics, Department of Physics, University of Erlangen-N\"{u}rnberg, Staudtstrasse 7, 91058 Erlangen, Germany}
\affiliation{Max Planck Institute for the Science of Light, Staudtstrasse 2, 91058
Erlangen, Germany}


\date{\today}

\begin{abstract}

We analyze the optical resonances of a dielectric sphere whose surface has been slightly deformed in an arbitrary way. Setting up a perturbation series up to second order, we derive both the frequency shifts and modified linewidths. Our theory is applicable, for example, to freely levitated liquid drops or solid spheres, which are deformed by thermal surface vibrations, centrifugal forces or arbitrary surface waves.
A dielectric sphere is effectively an open system whose description requires the introduction of non-Hermitian operators characterized by complex eigenvalues and not normalizable eigenfunctions. We avoid these difficulties  using the Kapur-Peierls formalism which enables us to extend the popular Rayleigh-Schr\"{o}dinger perturbation theory to the case of electromagnetic Debye's potentials describing the light fields inside and outside the near-spherical dielectric object. We find analytical formulas, valid within certain limits, for the deformation-induced first- and second-order corrections to the central frequency and bandwidth of a resonance. As an application of our method, we compare our results with preexisting ones finding full agreement.
\end{abstract}

\maketitle

\section{Introduction}\label{Introduction}

In this paper we address the problem of determining the optical resonances  of slightly deformed  dielectric spheres. Inside an almost spherical dielectric body embedded in vacuum or air, light is  confined by near-total internal reflection  and  propagates with little attenuation  along the inner surface of the body. This form  of propagation is denoted as whispering gallery modes (WGMs), which are typically characterized  by a high quality factor $Q$ \cite{Oraevsky}. For a perfect (ideal) dielectric sphere in air or vacuum, the predicted $Q$ can easily exceed  $10^{20}$ at optical frequencies. However, several physical processes (amongst which scattering from surface roughness can be the most prominent one), limit the effective value of $Q$ to less than $10^6$ \cite{PhysRevA.41.5199,LPOR:LPOR200910016}. Our goal is to develop a perturbation theory that allows us to calculate the  $Q$ factor of the optical resonances of   dielectric spheres whose surface is slightly deformed by different physical processes.

%
\begin{figure}[!ht]
\centerline{\includegraphics[scale=3,clip=false,width=.6\columnwidth,trim = 0 0 0 0]{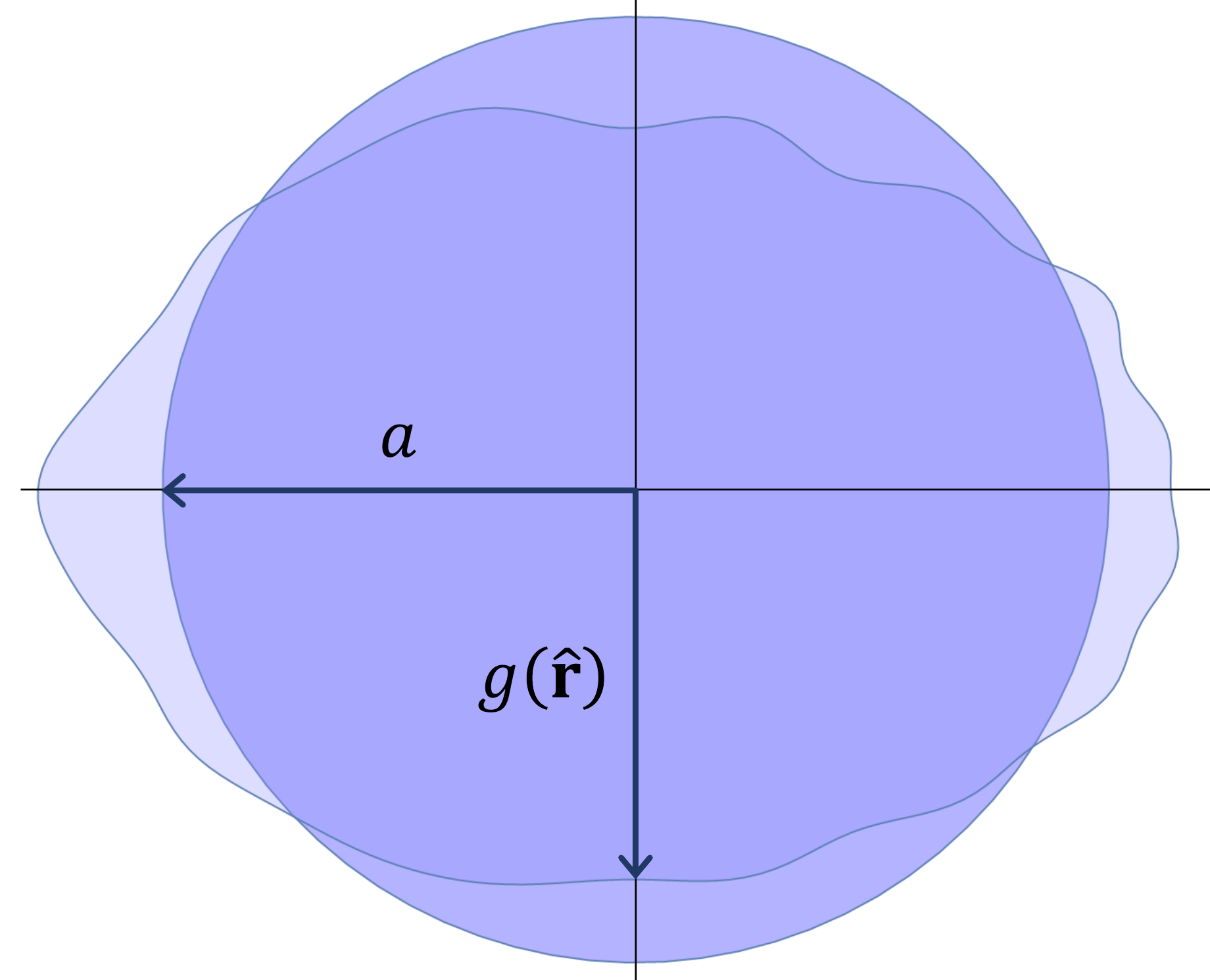}}
\caption{\label{fig0}  Cartoon-like representation of the cross-section of a  dielectric sphere with its deformations. The non deformed sphere is represented by a dark-blue disk of radius $a$.
The corrugated curve represents a generic quasi-spherical corrugated surface and it is characterized by the surface profile function $g(\hbr)$ (see sec. \ref{wave2} for details).}
\end{figure}
%

The study of light interacting with spherical or near spherical dielectric bodies dates back to Aristotle who first described (although incorrectly) the rainbow as due to light reflection from raindrops \cite{Rainbow}. In much more recent times microscopic glass spheres have been  widely used as passive and active optical resonators in linear and nonlinear optics regimes for numerous  physical, chemical, and biological applications (see, e.g., \cite{Foreman:15,Lin:17} and references therein). Lately, dielectric optical resonators of many diverse shapes have been regarded as optomechanical systems \cite{RevModPhys.86.1391,CavOpt}. Even more recently, optomechanical devices consisting of drops of various liquid materials have been proposed and demonstrated \cite{QIAN486,Dahan:16,PhysRevA.96.063842}. In these devices the near-spherical free surface of the drop provides for both the optical and the mechanical resonators. As an example thereof,  we have suggested the use of millimeter-scale drops of superfluid He magnetically levitated in vacuum as a novel type of optomechanical device \cite{PhysRevA.96.063842}. The surface of a levitated drop may differ from a perfect sphere for several reasons, as shown in Fig. \ref{fig0}. For example, a rotating liquid drop is squeezed along the axis of rotation and takes the form of an oblate spheroid. On top of this,  thermally excited capillary waves (ripplons) will result in corrugations upon the droplet's surface.

All these optical and optomechanical devices are describable as \emph{open systems}, that is physical systems that leak energy via the coupling with an external environment \cite{Gardiner}.
%
%
%
The mathematical description of either classical or quantum open systems requires the use of non Hermitean operators, which are characterized by complex-valued eigenvalues \cite{PhysRevC.6.114,Rotter2009,Brody2014}.
%
%
%
One important challenge with non Hermitean operators is that they may not possess a set of orthonormal eigenfunctions. This implies that the familiar Rayleigh-Schr\"{o}dinger perturbation theory is no longer applicable and different methods must be used.

Amongst these methods,  the quasi-stationary states approach and the Kapur-Peierls formalism are quite popular \cite{Kukulin}. Quasi-stationary (Gamow or Siegert functions \cite{Gamow1928,PhysRev.56.750}) states are solutions of a wave equation with  purely outgoing boundary conditions and can be used to build a perturbation theory called ``resonant-state expansion (RSE)'' \cite{PhysRevC.47.1903}. In optics, the RSE technique has been put forward in \cite{Muljarov10} and successfully applied to three-dimensional dielectric resonators in  \cite{PhysRevA.90.013834}. However, basically the same method was already used in \cite{PhysRevA.41.5187} to find optical resonances in microdroplets within first-order perturbation theory. The main problem with quasi-stationary states is that they are not orthonormal in the conventional sense and the standard normalization integral diverges \cite{Bohm89,Madrid12}. 

Conversely, Kapur-Peierls theory is not affected by these normalization problems  \cite{Kapur277,peierls1948} and automatically furnishes a biorthogonal complete set of functions suitable for  use in perturbation theory. This formalism was originally developed in the context of nuclear scattering theory and was recently applied to the study of the resonances of one- and two-dimensional open optical systems \cite{PhysRevA.67.013805,Viviescas2004}.

In this work we  use Kapur-Peierls formalism to develop, for the first time, a perturbation theory of optical resonances of three-dimensional open optical systems (near-spherical dielectric bodies), correct up to and including second-order terms.
We  find analytic formulas for the characteristic values (complex wave numbers) of these resonances and we  apply our theory to dielectric spheres with various deformations.

The application we have in mind is a situation in which the wavelength is much smaller than the sphere's radius (e.g. $100$ or $1000$ times). In this case, which is of great experimental significance, the use of numerical techniques (like the ones routinely used in commercially available  finite element method (FEM) solvers) becomes very challenging if not prohibitive. For this reason, we  do not present comparison to FEM results in the present manuscript.  However, we  compare our results with the analytical predictions (limited to first-order perturbation theory), of previous works and  find complete agreement.

The paper is organized as follows. In Sec. II we briefly describe what we regard as ``the unperturbed problem'', namely the determination of the optical resonances of a dielectric sphere using the formalism of Debye potentials and scattering theory. Then, in Sec. III we furnish a review of the Kapur-Peierls formalism, which sets the basis for the remainder.  In Sec. IV we apply this formalism to develop a perturbation theory for the Debye potentials. In Sec. V we describe in detail both the deterministic and random deformations of the initially spherical dielectric body. Then, in Sec. VI we use Rayleigh-Schr\"{o}dinger perturbation theory to achieve the main goal of this work, namely finding  the optical resonances of slightly deformed dielectric spheres. Finally, in Sec. VII we draw some conclusions.

\section{Resonances  of a dielectric sphere}\label{Dielectric sphere}

The mathematical problem of the interaction of electromagnetic waves with dielectric spheres is more than one century old and represents a vast literature. The standard reference is still Stratton's classic book \cite{Stratton}. However, a more modern and thorough exposition can be found in \cite{Grandy}. In this section we briefly review  the so-called Debye potentials approach and establish the basic notation that we shall use throughout this work.

\subsection{Setting the problem}\label{IIA}

Consider a sphere of radius $a$ made of a homogeneous  isotropic dielectric medium (medium $1$)  
surrounded by air or vacuum (medium $2$).
 We use SI units with electric permittivity $\veo$, magnetic permeability $\mu_0$ and speed of light $c = 1/(\veo \mu_0)^{1/2}$ in vacuum.
Let $\vE_1,\vB_1,\vD_1,\vH_1$ and $\vE_2,\vB_2,\vD_2,\vH_2$ denote the electromagnetic fields in medium $1$ and medium $2$, respectively. For our purposes it is sufficient to presume that all fields vary as $\exp(- i \omega t)$, where $\omega = k c$, $k$ being the wave number of light in vacuum. These fields obey the Maxwell equations
%
%
%
%
%
%
%
%
%
%
\begin{subequations}\label{Maxwell}
\begin{align}
\bnabla \cdot \vD_j &\; =0, \label{MaxwellA}\\
\bnabla \cdot \vB_j &\; =0, \label{MaxwellB} \\
 i \omega \vD_j + \bnabla \times \vH_j &\; =0, \label{MaxwellC} \\
- i \omega  \vB_j + \bnabla \times \vE_j &\; =0, \label{MaxwellD}
\end{align}
\end{subequations}
(here and hereafter $j=1,2$, unless stated otherwise) and the constitutive equations
\begin{align}\label{constitutive}
\vD_j  = \varepsilon_j \veo \vE_j, \qquad \text{and} \qquad \vB_j  = \mu_j \mu_0 \vH_j,
\end{align}
with $\mu_1 = \mu_2=1$ (we assume that both media are nonmagnetic) and  $\varepsilon_1=n_1^2, \; \varepsilon_2=n_2^2$, where $n_1 >1$ is the real-valued refractive index of medium $1$ and $n_2=1$ is the  refractive index of air or vacuum. The assumption that the dielectric is nonmagnetic implies that there is no physical difference between the magnetic strength $\vH$ and the magnetic induction $\vB$, so in the remainder we shall consider $\vB$ as the independent field.

Following \cite{Zangwill}, we express the solutions of the set of equations \eqref{Maxwell} in terms of the transverse electric (TE) and transverse magnetic (TM)  Debye \emph{scalar potentials} $\Psi_j(\br)$ and $\Phi_j(\br)$, respectively, as follows:
\begin{equation}\label{TE}
\begin{split}
\vE_j^\text{TE} = & \;  i k \bnabla \times ( \br \Psi_j ) , \\[6pt]
 c \vB_j^\text{TE} = & \; \bnabla \times \bigl[ \bnabla \times ( \br \Psi_j ) \bigr],
\end{split}
\end{equation}
%
%
%
%
%
%
%
and
\begin{equation}\label{TM}
\begin{split}
\vE_j^\text{TM} = & \;  \frac{i}{n_j^2}  \bnabla \times \bigl[ \bnabla \times ( \br  \Phi_j ) \bigr], \\[6pt]
c \vB_j^\text{TM} = & \; k \bnabla \times ( \br \Phi_j ).
\end{split}
\end{equation}
%
%
%
%
%
Equations \eqref{Maxwell} and \eqref{constitutive} are automatically satisfied by the fields \eqref{TE} and \eqref{TM} when the  Debye potentials obey the scalar Helmholtz equation
%
%
%
%
%
\begin{align}\label{Debye}
\nabla^2 U + k^2 n^2_j U =0 ,
\end{align}
where $U$ denotes either $\Psi_j$ or $\Phi_j$.
%
%
%
%
%
%
%
%
This equation  must be completed by the interface conditions which  require the continuity of  tangential components of $\vE$ and $\vH$ (or, $\vB$)  across the surface of the sphere \cite{Jackson}, that is 
\begin{equation}\label{boundaryTan}
\begin{split}
\hbr \times   \left( \vE_2 - \vE_1 \right) \bigr|_{r\, = \,a} = & \; 0, \\[4pt]
\hbr \times   \left( \vB_2 - \vB_1 \right) \bigr|_{r \,= \, a} = & \; 0,
\end{split}
\end{equation}
%
%
%
%
%
%
where $r= \abs{\br}$ and $\hbr = \br/r$.

Because of the  symmetry of the problem imposed by \eqref{boundaryTan},  it is convenient to solve the Helmholtz equation \eqref{Debye} in spherical coordinates $(r,\theta,\phi)$. Following \cite{Zangwill} we rewrite the Laplace operator $\nabla^2$ as
\begin{align}\label{E20}
\nabla^2 U = \frac{1}{r} \frac{\partial^2}{\partial r^2} \bigl( r U  \bigr)- \frac{\hat{L}^2}{r^2} \, U ,
\end{align}
where $\hat{L}^2 \equiv \mathbf{\hat{L}} \cdot \mathbf{\hat{L}}$ with $\mathbf{\hat{L}} \equiv - i \, \br \times \bnabla$. Now, we look for solutions of \eqref{Debye} of the form
\begin{equation}\label{Sepa}
\begin{split}
\Psi_j(r,\theta,\phi) =    \frac{u_j(r)}{r} \, Y_{lm}(\theta,\phi), \\
\Phi_j(r,\theta,\phi) =   \frac{ v_j(r)}{r} \, Y_{lm}(\theta,\phi),
\end{split}
\end{equation}
where $Y_{lm}(\theta,\phi)$ are the standard spherical harmonics \cite{Jackson} satisfying $\hat{L}^2 Y_{lm} = l(l+1)Y_{lm}$, and $ u_j(r)$, $ v_j(r)$ denotes the \emph{reduced radial Debye potentials}. Substituting \eqref{Sepa} into \eqref{Debye} and using \eqref{E20}, we obtain the ordinary differential equation
%
%
%
%
%
\begin{align}\label{eqPsi}
- \psi_j''(r) + \left[ \frac{l(l+1)}{r^2} - k^2 n_j^2 \right] \psi_j(r) =0,
\end{align}
where  $\psi_j = u_j$ for TE polarization, $\psi_j = v_j$ for TM polarization and $\psi_j'' \equiv \di ^2 \psi_j / \di r^2$. This equation must be supplied with  the interface conditions for the reduced radial potentials $ \psi_j(r)$. Substituting \eqref{TE} and \eqref{TM} into \eqref{boundaryTan} and using \eqref{Sepa} we obtain
\begin{align}\label{BC}
\psi_1(a)   =   \psi_2(a), \qquad \psi_1'(a) &  =  p \, \psi_2'(a),
\end{align}
where $\psi_j' \equiv \di \psi_j/ \di r$ and here and hereafter $p =1$ for TE polarization and $p = n_1^2/n_2^2$ for TM polarization. We remark that in the literature equation \eqref{eqPsi} is often written in a ``quantum-like'' form as
\begin{align}\label{eqPsi2}
- \psi_j''(r) + \left[\frac{l(l+1)}{r^2} + V_j \right]\psi_j(r) = E \, \psi_j(r),
\end{align}
where $V_j = k^2 \left( 1 - n_j^2\right)$ and $E = k^2$ (see, e.g., \cite{Zangwill} and \cite{Johnson:93}). We shall exploit this quantum-classical analogy in the next section.

\subsection{Scattering solutions}\label{IIB}

The  general solution of \eqref{eqPsi} can be written  as
\begin{align}\label{GenSol1}
\psi_j(r) = C_1  \, r \, j_l(n_j k r) + C_2 \, r \, y_l(n_j k r),
\end{align}
%
%
%
%
%
where $j_l(z)$ and  $y_l(z)$ are spherical Bessel functions of the first and second kind, respectively \cite{Newton}. Using the spherical Hankel functions 
$h^{(1)}_l(z) = j_l(z) + i \, y_l(z)$ and  $h^{(2)}_l(z)= j_l(z) - i \, y_l(z)$, we can rewrite \eqref{GenSol1} as
\begin{align}\label{GenSol2}
\psi_j(r) = C_3  \, r \, h^{(1)}_l(n_j k r) + C_4 \, r \,  h^{(2)}_l(n_j k r),
\end{align}
where $C_3 = (C_1 - i C_2)/2$ and $C_4 = (C_1 + i C_2)/2$.
%
%
%
%
%
Since $j_l(z) \sim z^l$ and $n_l(z) \sim 1/z^{l+1}$ for $z \to 0$, while $h^{(1)}_l(z) \sim (-i)^{l+1} {e^{i z}}/{z}$ and $ h^{(2)}_l(z) \sim i^{l+1} {e^{-i z}}/{z}$ for $z \to \infty$,  the \emph{everywhere regular} solutions to \eqref{eqPsi} are:
\begin{align}\label{Radial}
\psi_1(r) = & \;  A_{ l} \, r \, j_l(n_1 k r) , &  r \leq a, \nonumber \\
 & \; \\[-8pt]
\psi_2(r) = & \;    I  \, r \,     h^{(2)}_l(k r) + S_l \, r \, h^{(1)}_l(k r), & r>a, \nonumber
\end{align}
where $I$ is the amplitude of the incident wave and $S_l$ that of the scattered wave with azimuthal index $l$. $A_l$ is the amplitude of the same wave inside the sphere.
%
Assuming only \emph{outgoing waves} means setting $I=0$. This choice leads to the so-called ``resonant-state'' formulation of scattering theory \cite{Muljarov10,PhysRevA.90.013834}. These states, also known in the quantum theory of scattering \cite{Kukulin} as decaying, meta-stable, Gamow \cite{Gamow1928}, or Siegert \cite{PhysRev.56.750} states, are nonphysical because they are not normalizable in the standard manner (that is, they are not square-integrable).
Here  we choose instead $I=1$, which means assuming an \emph{incident wave} of unit amplitude.

%
Substituting \eqref{Radial}  into \eqref{BC} we determine the interior wave amplitude
\begin{align}\label{ampA}
A_l(k) = \frac{2 \, i p}{k a}\frac{1}
{f_l(k a)},
\end{align}
and the scattering amplitude
\begin{align}\label{ScatMat}
S_l(k) = -\frac{f_l(-ka)}{f_l(ka)},
\end{align}
where we have defined the Jost function \cite{GIAMBIAGI195939},
\begin{multline}\label{Jost}
f_l(z) = \\ p \, j_l(n_1 z) \bigl[ z \, h^{(1)}_l(z) \bigr]' - h^{(1)}_l(z)\bigl[(n_1 z) j_l(n_1 z) \bigr]',
\end{multline}
with the prime symbol $(')$ denoting the derivative with respect to the argument of the function (e.g., $[f(x) g(x) ]' = (\di f/ \di x) g(x) +f(x)(\di g/ \di x) $).
Using \eqref{AnCont} it is straightforward to show that for $k$ real, $f_l(-k a)=f_l^*(k a)$ and we can write
\begin{align}\label{ScatMat2}
S_l(k) =   \exp \left[ 2 i \delta_l(k) \right],
\end{align}
where $\delta_l(k)$ denotes the phase shift of the scattered wave \cite{Grandy}. In the absence of the dielectric sphere $n_1=n_2 =1$ and evidently scattering does not occur. In this case the equations above give $\delta_l=0$, $S_l=1$ and $A_l=2$.

\subsection{Resonances and Q-factors}\label{Q-factor}

\begin{figure}[!hb]
\centerline{\includegraphics[scale=3,clip=false,width=1\columnwidth,trim = 0 0 0 0]{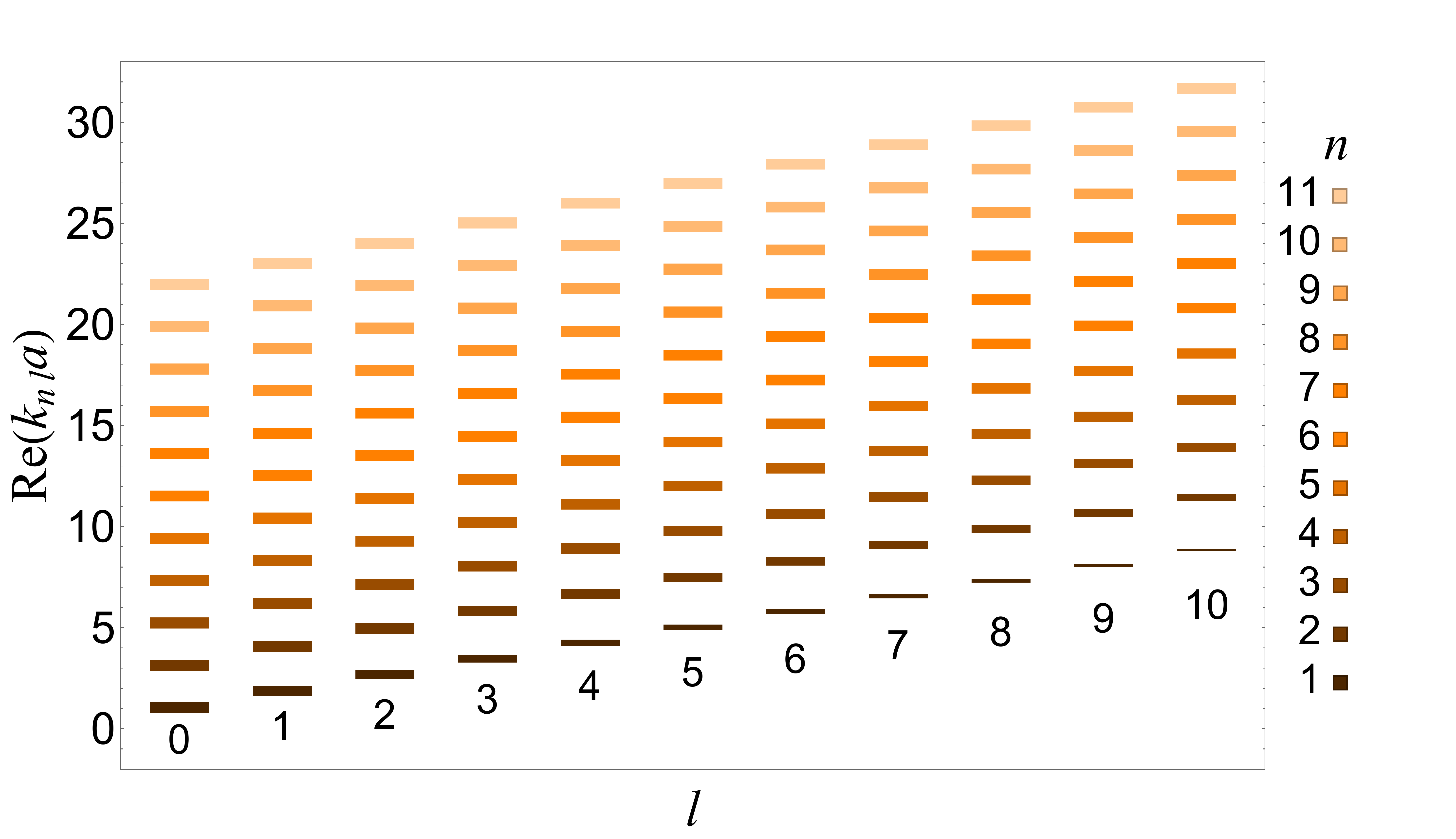}}
\caption{\label{fig6}  Spectrum of the TE modes of a dielectric sphere of radius $a$ and refractive index $n_1=1.5$. The values of $k_{nl}$ for $1 \leq n \leq 11$ and  $0 \leq l \leq 10$ are shown as orange bands. The vertical position of the center of each band  is equal to $\re (k_{nl} a)$ and
the thickness equal to $\im (k_{nl}a)$. For the first radial mode $n=1$ (darkest orange bands) the imaginary part of $k_{nl}$ quickly decreases as $l$ increases from left to right, while it decreases  slowly for $n>1$ radial modes.
 Each mode characterized by the pair of radial and azimuthal numbers $(n,l)$ is $2l+1$ times degenerate (see sec. \ref{PertDebye}).}
\end{figure}

In equations \eqref{ampA} and \eqref{ScatMat}  $k$ is the \emph{real-valued} wave number of the ingoing wave. 
However, the resonances of the sphere are associated with the poles of the analytical continuation of   $S_l(k)$ into the entire complex plane: $ k \in \mathbb{R} \to k = k' + i k''\in \mathbb{C}$, where here and hereafter $k' = \re k$ and $k'' = \im k$.
The continuation of $S_l(k)$ is meromorphic, that is analytic except at its poles. The latter are characterized by $ \im k < 0$ and coincide with  the roots of  the transcendental equation
\begin{align}\label{Poles}
f_l(k a) =  0.
\end{align}
This equation, where $l$ is a \textbf{fixed} number, has a denumerably infinite set of solutions denoted $\{ k_{1l} , k_{2l},  \ldots, k_{nl}, \ldots\}$ whose determination is detailed in appendix \ref{A}. From \eqref{AnCont} it follows that  $f_l^*(z) =  f_l(-z^*)$, that is the resonance poles are located in the complex $k$-plane in pairs symmetric with respect to the imaginary axis. Therefore, if $k_{nl}$ is a solution of \eqref{Poles}, then $-k^*_{nl}$ is also a solution. We label the poles with $\re k <0$ by the negative index $-n$, so that $k_{-n l} = - k_{nl}^*$. A ``central'' pole labeled with $n=0$ and characterized by  $\re k_{0l} = 0$, $\im k_{0l} < 0$, exists only for $l$ odd (even) and TE (TM) polarization. A portion of the spectrum of TE resonances of a dielectric sphere with refractive index $n_1=1.5$, is shown in Fig. \ref{fig6}. 
%
%

Each resonance $k = k_{1l}, k_{2l}, \dots$, can be characterized by the {quality factor} $Q$ defined by
\begin{align}\label{Q}
Q(k',k'') \equiv - \frac{1}{2}\frac{\re k}{\im k} = - \frac{1}{2} \frac{k'}{k''}.
\end{align}
 From this equation it follows that
\begin{align}\label{Euler}
\frac{\partial Q}{\partial k''} = 2 \, Q   \frac{\partial Q}{\partial k'} .
\end{align}
This means that $Q$ is more sensitive to variations of losses ($\propto k''$) than of frequency ($\propto  k'$), by a factor  $ 2 Q$. This is why also a tiny perturbation of $k''$ may cause a relevant variation of $Q$.
This feature is relevant  for the estimation of the variation of $Q$ due to a small perturbation of the shape of the dielectric sphere.

The quality factor depends dramatically upon the value of $l$. For example, solving equation \eqref{Poles} numerically for a $\Heq$ sphere with refractive index $n_1 \approx 1.03$ (superfluid He),  $p=1$ (TE polarization), $l=4000$ and $l=1000$, we have found  $k_{1,4000} a \approx 4000/n_1 - i \,( 2 \times 10^{-10})$ and $k_{1,1000}a \approx 1000/n_1 - i \, (1 \times 10^{-1})$, respectively, where we have chosen in both cases the first resonance labeled by $n=1$. These values yield
\begin{align}
Q(4000) = - \frac{1}{2} \frac{\re(k_{1,4000} a)}{\im(k_{1,4000} a)} \approx  10^{13},
\end{align}
for $l=4000$, and
\begin{align}
Q(1000) = - \frac{1}{2} \frac{\re(k_{1,1000} a)}{\im(k_{1,1000 } a)} \approx 5 \times 10^{3} ,
\end{align}
for $l=1000$. Thus, although $l$ changes only by a factor of $4$, the corresponding $Q$ changes by about $9$ orders of magnitude. This huge variation in $Q$ is largely determined by the imaginary parts of the resonances, because
\begin{align}
\frac{Q(4000)}{Q(1000)} = & \; \frac{\re(k_{1,4000} a)}{\re(k_{1,1000}a)} \times \frac{\im(k_{1,1000} a)}{\im(k_{1,4000 }a)} \nonumber \\[6pt]
 \approx & \; \frac{4000}{1000} \times ( 2 \times 10^9 ).
\end{align}

\section{Kapur-Peierls formalism}\label{KP}

In the previous section we have presented the standard theory of scattering from a piecewise constant spherically symmetric potential (dielectric sphere) and we have written the equation \eqref{Poles} determining the resonances of the system \cite{Newton}. This approach, based on the continuous (with respect to $k$) set of functions \eqref{Radial}, is not very convenient for perturbation theory where it is desirable to deal with a denumerable set (a basis) of normalizable functions. The Kapur-Peierls (KP) formalism, originally developed in the context of nuclear physics \cite{Kapur277} and recently adapted to optical resonator theory \cite{PhysRevE.67.026215,Viviescas2004}, naturally yields a  complete set of biorthogonal functions \cite{PhysRevA.7.1288}.

\subsection{Preliminaries on Kapur-Peierls formalism}\label{KPa}

Before starting our discussion, it is useful to briefly outline the general approach of KP perturbation theory. In the standard quantum mechanics Rayleigh-Schr\"{o}dinger time-independent perturbation theory, one first finds the full set of eigenstates of the  Hamiltonian of the unperturbed system. Afterwards, the perturbative corrections to any one eigenstate can be expressed generically as sums over these eigenstates. In KP  perturbation theory, the setting is slightly changed: One first solves an \emph{auxiliary} eigenproblem whose eigenvalues $\lambda(k)$ are functions of a continuous parameter, the complex scattering frequency (here represented by the complex wavenumber $k$). One then determines the discrete set of resonances in $k$ by imposing $\lambda(k) = k^2$. Finally, the perturbative correction for a given resonance is obtained by summing over the previously obtained set of eigenstates that belongs to the resonance's particular value of $k$. This makes the whole procedure more involved than Rayleigh-Schr\"{o}dinger theory, since for each resonance we are dealing with a different set of infinitely many eigenstates (which are still loosely related to the whole set of resonances, but not identical to those).

 Kapur-Peierls dispersion theory is well known within nuclear physics \cite{RevModPhys.31.893}. However, this formalism is much less known in the optics community. A useful purpose may therefore be served by shortly reviewing the Kapur-Peierls  approach to scattering theory \cite{Johnson}. As in the previous section, we consider again the scattering of a scalar wave  {(any of the two Debye potentials)} by a dielectric sphere; this simple example illustrates the main features of the theory and provides for the Kapur-Peierls eigenvalues and eigenfunctions characterizing the ``unperturbed problem''. When the scatterer is not perfectly spherical the simple theory presented in this section is no longer applicable and the use of perturbation theory becomes necessary. This will be presented in the next section.

We begin by rewriting \eqref{eqPsi} as
\begin{align}\label{Compact}
\bigl( \hD_j - k^2 \bigr) \psi_j(r) = 0, \qquad (j=1,2),
\end{align}
where we have defined the differential operator
\begin{align}\label{OpD}
 \hD_j \equiv \frac{1}{n_j^2} \left[ -\frac{\di^2}{\di r^2} + \frac{l(l+1)}{r^2}  \right],
\end{align}
associated with the boundary conditions \eqref{BC} that we rewrite as:
%
%
%
%
%
\begin{align}\label{BCOpD}
\frac{\psi_1'(a)}{\psi_1(a)}  = p \, \frac{\psi_2'(a)}{\psi_2(a)} ,
\end{align}
where $p =1$ for TE polarization and $p = n_1^2/n_2^2$ for  TM polarization. We know from the previous section that the solution of \eqref{Compact}  can be written for $r>a$ as
\begin{align}\label{Psi2}
\psi_2(r) = & \;    I  \, r \,     h^{(2)}_l(k r) + S_l \, r \, h^{(1)}_l(k r),
\end{align}
which implies,
\begin{align}\label{Psi2der}
\psi_2'(r) = & \;    I  \, \bigl[(kr) \,     h^{(2)}_l(k r) \bigr]' + S_l \,  \bigl[(kr)\, h^{(1)}_l(k r)\bigr]',
\end{align}
where the prime symbol $(')$ denotes the derivative with respect to the argument of the function.  

Kapur-Peierls theory is based upon the observation that
using \eqref{Psi2} and \eqref{Psi2der} we can express $S_l$ and $I$ via $\psi_2(a)$ and $\psi_2'(a)$ to obtain
\begin{align}\label{ratioSI}
\frac{S_l}{I} = -\frac{h_l^{(2)}(ka)}{h_l^{(1)}(ka)} \, \frac{\psi_2'(a) + c_{\,l}(-k a)\psi_2(a)}{\psi_2'(a) - c_{\,l}(k a)\psi_2(a)},
\end{align}
where
\begin{align}\label{Cl}
c_{\,l}(k a) \equiv \frac{1}{a} \, \frac{ \bigl[(ka)\, h^{(1)}_l(k a)\bigr]'}{h^{(1)}_l(k a)}.
\end{align}
In Secs. \ref{IIB} and \ref{Q-factor} we have shown that the poles of the analytic continuation of $S_l(k)$, with $k = k' + i k''$, determines the resonances of the systems. From \eqref{ratioSI} it follows that these poles occur when the denominator vanishes, that is when $\psi_2'(a) - c_{\,l}(k a)\psi_2(a)=0$. Evidently, this happens when there is no incident wave, that is $I=0$ and the ratio ${S_l}/{I}$ becomes singular.
Using the boundary conditions \eqref{BCOpD} we can transform the  relation $\psi_2'(a) - c_{\,l}(k a)\psi_2(a)=0$ into the equivalent one,
\begin{align}\label{EqPolesKP}
\psi_1'(a) - p \, c_{\,l}(k a)\psi_1(a)=0.
\end{align}
%
%
%
%
%
%
This implies that we can determine the resonances of the system by knowing the solutions $\psi_1(r) $ of the \emph{interior} problem $\bigl( \hD_1 - k^2 \bigr) \psi_1(r) = 0$ with boundary conditions \eqref{EqPolesKP}.
We shall give a constructive proof of this statement in  subsection \ref{KPa3} by deriving the so-called \emph{dispersion formula} for the scattering amplitude $S_l(k)$.  However, first we need to prove some basic results. 

\subsection{The Kapur-Peierls eigenfunctions}\label{KPa2}

Let us consider the auxiliary eigenvalue problem
\begin{align}\label{Aux1}
\bigl( \hD_1 - \lambda_{nl}(k) \bigr) \phi_{nl}(k,r) & \; = 0, \qquad r \leq a,
\end{align}
with boundary conditions
\begin{align}\label{AuxBC}
\phi_{nl}(k,0) = 0, \quad \phi_{nl}'(k,a) - p \,c_{\,l}(ka)\phi_{nl}(k,a) = 0,
\end{align}
 where $n$ is a discrete numerical index, $\phi_{nl}(k,r)$ are the so-called Kapur-Peierls (right) eigenfunctions with $\phi_{nl}'(k,a) \equiv [\di \phi_{nl}(k,r)/ \di r]_{r=a}$, and $c_{\,l}(ka)$ is given by \eqref{Cl}. The (right) eigenvalues $\lambda_{nl}(k)$ depend on the parameter $k$ via the boundary conditions \eqref{AuxBC}. Here and hereafter $k$  must be regarded as a fixed constant, the \textit{same} for all eigenvalues $\lambda_{1l}(k), \lambda_{2l}(k), \ldots$, which are complex numbers on account of the boundary condition \eqref{AuxBC}.
The normalized solutions of \eqref{Aux1} are
\begin{align}\label{KPfun1}
\phi_{nl}(k,r) = \frac{1}{\sqrt{Z_{nl}}} \; r \, j_l \left( n_1  q_{nl}  r\right),
\end{align}
where $q_{nl} =\sqrt{\lambda_{nl}(k)}$
%
%
%
%
and
\begin{align}\label{Norm}
Z_{nl} = \frac{a^3 }{2}\Bigl[j_{l}^{\,2}( n_1 q_{nl} a) - j_{l-1}( n_1 q_{nl}  a) j_{l+1}( n_1 q_{nl}  a) \Bigl].
\end{align}
The eigenvalues are given by $\lambda_{nl}(k) = z_{nl}^2/(n_1 a)^2$, where $\{z_{1l}, z_{2l}, \ldots, z_{nl}, \ldots \}$, are  the complex roots of the $k$-dependent transcendental equation $F_l(z,k a) = 0$, where
%
%
%
%
%
\begin{align}\label{EigenKP}
F_l(z,w) = p \, j_l(z) \bigl[ w \, h^{(1)}_l(w) \bigr]' - h^{(1)}_l(w)\bigl[z \, j_l( z) \bigr]' .
\end{align}
From \eqref{diff} it follows that if $z_{nl}$ is a solution of \eqref{EigenKP}, then $-z_{nl}$ is also a solution and both $z_{nl}$  and $-z_{nl}$  yield the same eigenvalue $\lambda_{nl}(k)$. Different values of $k$ produce different eigenvalues; typically $\lambda_{nl}(k) \neq \lambda_{nl}(k')$ for $k \neq k'$.

The operator defined by \eqref{Aux1} and \eqref{AuxBC} is \emph{not} self-adjoint because $c_{\,l}(k a)$ is a complex number. This implies that there exist left eigenfunctions $\tphi_n(k,r)$ and left  eigenvalues $\tla_{nl}(k)$  defined by the so-called adjoint equation
\begin{align}\label{Aux1adj}
\bigl( \hD_1 - \tla_{nl}(k) \bigr) \tphi_{nl}(k,r) & \; = 0, \qquad r \leq a,
\end{align}
and the adjoint boundary conditions
\begin{align}\label{AuxBCadj}
\tphi_{nl}(k,0) = 0, \quad \tphi_{nl}'(k,a) - p \,c_{\,l}^*(ka)\tphi_{nl}(k,a) = 0.
\end{align}
It is not difficult to show  that $\tla_{nl}(k) = \lambda_{nl}(k) = \left[ \lambda_{nl}(-k^*) \right]^*$ and $\tphi_{nl}(k,r) =  \phi_{nl}^*(k,r) = \phi_{nl}(-k^*,r)$ \cite{PhysRevA.7.1288}.
Moreover,  our  normalization \eqref{Norm}  yields
\begin{align}\label{Ortho}
\int_0^a \tphi_{{n'l}}^*(k,r) \phi_{nl}(k,r) \di r = & \; \int_0^a \phi_{n'l}(k,r) \phi_{nl}(k,r) \di r  \nonumber \\[6pt]
= & \;\delta_{n  n'}.
\end{align}
This equation shows that the normalized Kapur-Peierls eigenfunctions $\phi_{nl}(k,r)$ belong to a  biorthogonal set of functions.

Typically the  functions $\phi_{nl}(k,r)$ form a complete set \cite{peierls1948,Fonda66}, that is
\begin{align}\label{Complete}
\sum_n \phi_{nl}(k,r) \tphi_{nl}^*(k,r')  = & \; \sum_n \phi_{nl}(k,r) \phi_{nl}(k,r')   \nonumber \\[6pt]
= & \; \delta\left( r - r' \right),
\end{align}
but usually this is not easy to prove (see, e.g., \cite{Dolph1961} for a discussion).
For our functions \eqref{KPfun1} we  have not been able to evaluate the left side of this equation analytically, but
numerical evaluation for some values of $l$ and $k$ confirmed the validity of \eqref{Complete}. Therefore, we assume without demonstration the completeness of the Kapur-Peierls functions \eqref{KPfun1}.

\subsection{The Kapur-Peierls dispersion formula}\label{KPa3}

From (\ref{BCOpD}-\ref{Psi2der}) it follows that the interior function $\psi_1(r)$ obeys the boundary conditions
\begin{align}\label{BCpsi1}
\psi_1'(a) -  p \,c_{\,l}(ka) \psi_1(a) =  I \frac{2 \,  p}{i \,  \xi_{\,l}(ka)},
\end{align}
where we have introduced the Riccati-Bessel functions $\xi_{\,l}(x) \equiv x \, h^{(1)}_l(x) $ and $\zeta_{\,l}(x) \equiv x \, h^{(2)}_l(x)$ \cite{Abramowitz}. These conditions reduce to \eqref{EqPolesKP} when no incident wave is present and $I=0$.
Consider then the auxiliary functions $\varphi_1(r)$ and $\varphi_2(r)$ defined by
\begin{align}\label{AuxFun}
\varphi_j(r) \equiv \psi_j(r) - X(r), \qquad (j=1,2),
\end{align}
where $X(r)$ is any function satisfying the constraint
\begin{align}\label{ConstraintX}
X'(a) - p \,c_{\,l}(ka) X(a) =  I \frac{2 \,  p}{i \, \xi_{\,l}(ka)}.
\end{align}
It is then evident that $\varphi_1(a)$ obeys the same boundary conditions \eqref{AuxBC} satisfied by the Kapur-Peierls functions, that is
\begin{align}\label{BCvarphi1}
\varphi_1'(a) -  p \,c_{\,l}(ka) \varphi_1(a) =  0.
\end{align}
Therefore, using \eqref{Complete} and \eqref{AuxFun} we can write
\begin{align}\label{Expansion}
\varphi_1(r) = \sum_n a_n \, \phi_{nl}(k,r),
\end{align}
where
%
%
%
%
%
%
%
%
%
%
%
%
\begin{align}\label{As}
a_n = & \; \int_0^a \phi_{nl} (k,r) \bigl[ \psi_1(r) - X(r)  \bigr]  \di r \nonumber \\[6pt]
%
%
\equiv & \; b_n - c_n.
\end{align}
From \eqref{Compact} and \eqref{OpD} and using $\psi_1(0)=0=\phi_{nl}(k,0)$, we obtain
\begin{align}\label{Bs}
b_n = & \; \int_0^a \phi_{nl} (k,r) \, \psi_1(r) \di r \nonumber \\[6pt]
= & \; \int_0^a \frac{\bigl( \hD_1 \phi_{nl} (k,r) \bigr) \psi_1(r) - \phi_{nl} (k,r)\bigl( \hD_1 \psi_1(r)\bigr)}{\lambda_{nl}(k) -k^2} \di r \nonumber \\[6pt]
= & \; \frac{1}{n_1^2} \, \frac{\phi_{nl}(k,a)}{\lambda_{nl}(k) -k^2} \, \bigl[\psi_1'(a) - p \, c_{\, l}(ka)  \psi_1(a) \bigr].
\end{align}
Subtracting $X(a)$ from both sides of the matching condition $\psi_2(a) = \psi_1(a)$ we obtain $\varphi_2(a) = \varphi_1(a)$. Using (\ref{Psi2},\ref{Expansion}) and \eqref{As} we can rewrite this equation as
\begin{multline}\label{Join}
\frac{1}{k} \left[ I \, \zeta_{\,l}(k a) + S_l \, \xi_{\,l}(k a) \right] - X(a) \\[4pt]
= \sum_n b_n \phi_{nl}(k,a) - \sum_n c_n \phi_{nl}(k,a).
\end{multline}
Substituting \eqref{Bs} into \eqref{Join} and using  \eqref{BCpsi1}, gives
\begin{align}\label{Join2}
\frac{1}{k} \bigl[ I \, \zeta_{\,l}(k a) \,+ \, & S_l \, \xi_{\,l}(k a) \bigr] \nonumber \\[4pt]
= & -I \, \frac{p}{n_1^2} \, \frac{2 i}{\xi_{\,l}(ka)} \sum_n \frac{\phi_{nl}(k,a)}{\lambda_{nl}(k) -k^2} \nonumber \\[4pt]
& + \Bigl[ X(a) - \sum_n c_n \phi_{nl}(k,a) \Bigr].
\end{align}
%
%
Since $X(r)$ is arbitrary and the condition \eqref{ConstraintX} involves both $X(a)$ and $X'(a)$, we  can always choose $X(r)$ such that $X(a) = \sum_n c_n \phi_{nl}(k,a)$ to cancel the last term in \eqref{Join2}, and $X'(a)$ in a manner that \eqref{ConstraintX} becomes an identity.
Then, solving \eqref{Join2} for $S_l$, we obtain
\begin{align}\label{Join3}
\frac{S_l}{I} = - \frac{\zeta_{\,l}(k a)}{\xi_{\,l}(k a)} \, \bigl[ 1 + 2 \, i\, k R_{\,l}(k) \bigr],
\end{align}
where
\begin{align}\label{Join4}
R_{\,l}(k) = \,\frac{p}{n_1^2} \, \frac{1}{\xi_{\,l}(k a)\zeta_{\,l}(k a)} \, \sum_n  \frac{\phi_{nl}^2(k,a)}{\lambda_{nl}(k) -k^2},
\end{align}
and $p =1$ for TE polarization and $p = n_1^2/n_2^2$ for  TM polarization.
It should be noticed that the sum in \eqref{Join4} is simply equal to $-1$ times the Green function $G_l(k,r',r)$ for the internal problem $r',r \leq a$, evaluated at $r'=r=a$ \cite{PhysRevA.7.1288}. We shall use this property later in Sec. \ref{PerTheory}.

Equations (\ref{Join3}-\ref{Join4}) are an example of what is usually called a ``dispersion formula'' in nuclear physics. They give an explicit expression of the scattering amplitude $S_l$ in terms of its  singularities (poles). In particular,  \eqref{Join4} provides for a practical recipe to find resonances: first we calculate the Kapur-Peierls eigenvalues $\lambda_{nl}(k)$ by solving (often numerically) the transcendental equation $F_l(n_1 a \sqrt{\lambda_{nl}(k)}, k a)=0$ to determine $\sqrt{\lambda_{nl}(k)}$. Then, we look for the roots of the fixed point equation
\begin{align}\label{Join5}
\sqrt{\lambda_{nl}(k)} = k.
\end{align}
It is understood that the only physically acceptable branch of the multi-valued function $\sqrt{\lambda_{nl}(k)}$ is the one with $\im \sqrt{\lambda_{nl}(k)} <0$.
It is evident that \eqref{Join5} reproduces the resonance equation \eqref{Poles}. To show this we must simply  substitute, consistently with \eqref{Join5}, $n_1 a \sqrt{\lambda_{nl}(k)}$ with $n_1 a \, k$ in  $F_l(n_1 a \sqrt{\lambda_{nl}(k)}, k a)=0$. This makes \eqref{EigenKP} coincident with \eqref{Poles}, that is $F_l(z,z) = f_l(z)$.

\begin{figure}[!ht]
\centerline{\includegraphics[scale=3,clip=false,width=1\columnwidth,trim = 0 0 0 0]{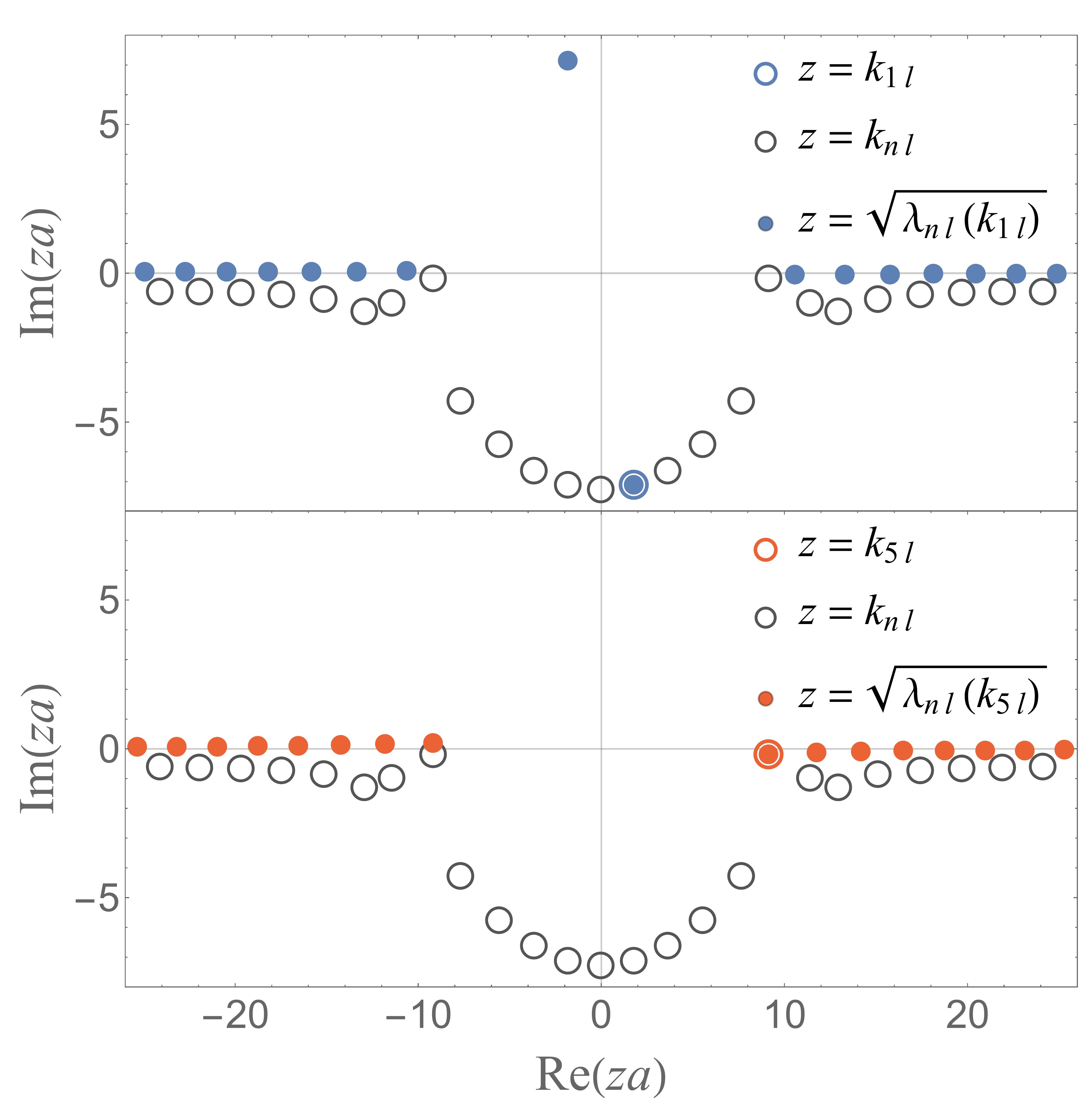}}
\caption{\label{fig4}   
Resonances $k$ of an unperturbed sphere, and discrete eigenvalues of the Kapur-Peierls equation for two different values of $k$.
The top plot displays the location of the roots $\{k_{nl} \}$, of \eqref{Poles} (open black circles) and \eqref{EigenKP} $\{ \sqrt{\lambda_{nl}(k_{1l})} \}$ (filled blue circles) for $k=k_{1l}$. The root denoted $k_{1l}$ ($k_{0l}$ is the central root with $\re k_{0l} =0$, $k_{1l}$ is the right nearest root with $\re k_{1l} >0$, $k_{-1l}$ is the left nearest root with $\re k_{-1l} <0$, et cetera) is indicated by a blue open circle. It is evident that $\sqrt{\lambda_{nl}(k_{1l})} = k_{1l}$ for only one value of $n$.
Similarly, the bottom plot displays the location of the roots of \eqref{Poles} (open black circles) and \eqref{EigenKP} $\{ \sqrt{\lambda_{nl}(k_{5l})} \}$ (filled red circles) for $k=k_{5l}$. The root denoted $k_{5l}$ is marked by a red open circle. Also here $\sqrt{\lambda_{nl}(k_{5l})} = k_{5l}$ for only one value of $n$.  In both plots the field has TM polarization, $n_1 = 1.5$ and $l=10$.  }
\end{figure}

We remark that for a fixed value of the index $n$, there may be several different solutions $k_{1l}, k_{2l}, \ldots, k_{sl},\ldots$, of \eqref{Join5}  such that $\lambda_{nl}(k_{sl}) = k_{sl}^2$. An example thereof is reported in \cite{PhysRevA.7.1288}. However, in our case we found via numerical evaluation of \eqref{Join5} that there is only one solution for fixed $n$; this is illustrated in Fig. \ref{fig4} for two particular cases. Therefore, in the remainder we choose the natural numeration of the resonances so that $s=n$ and $\lambda_{nl}(k_{nl}) = k_{nl}^2$.
%

%

%
%
%
%


\section{Perturbation theory for the Debye potentials}\label{PertDebye}

In the previous section we have described the Kapur-Peierls formalism. This yields a biorthogonal and complete set of basis functions defined in the interior region of the dielectric sphere. The goal of this section is  to develop a perturbation theory for the Helmholtz equation  \eqref{Debye} using these functions.

\subsection{Description of the deformations of the surface of a dielectric sphere}\label{wave2}


We assume that the sphere's free surface can be described in spherical coordinates $(r,\theta,\phi)\equiv (r,\hbr)$ by the equation $r - g(\hbr)=0$, where
\begin{align}\label{cavEq}
g(\hbr) \equiv a + a  h(\hbr),
\end{align}
is the surface profile function and $a  \abs{h(\hbr)}$ describes the distance, in the direction $\hbr$, of the \emph{deformed} sphere surface from a \emph{reference} unperturbed sphere of radius $a$. We suppose that for a given  fixed direction $\hbr$, the equation $r - g(\hbr)=0$ has only one solution. By definition, for a perfect sphere of radius $a$ the profile function is   constant, namely $g(\hbr) =a$ and $h(\hbr) =0$. Conversely,  the surface profile function of the deformed sphere is effectively determined by
\begin{align}\label{ThermalH}
h(\hbr) =  \sum_{L=2}^\infty \sum_{M=-L}^L h_{LM} Y_{LM}(\hbr),
\end{align}
where
\begin{align}\label{ThermalH2}
 h_{LM} = \int\limits_0^{2 \pi} \di \phi  \int\limits_0^{\pi}\di \theta  \sin \theta \,  Y_{LM}^*(\hbr)\, h(\hbr) .
\end{align}

The relative permittivity $\varepsilon_r(\br)$ of the deformed sphere can be evidently described by the piecewise constant function
\begin{align}\label{epsilonR}
\ver = n_1^2 H\bigl(g(\hbr) - r \bigr) + n_2^2 H\bigl( r - g(\hbr) \bigr),
\end{align}
where  $H(x)$ denotes the Heaviside step function \cite{Hfunction}.
For a perfect sphere  of radius $a$ we define $\varepsilon_r(\br)\equiv \varepsilon_r^{(0)}(r) =  n_1^2 \, H(a-r) + n_2^2H (r-a)$. From $ H(x) + H(-x)=1$ it follows that we can rewrite $\varepsilon_r(\br)$ as the sum of the unperturbed permittivity $\varepsilon_r^{(0)}(r)$ and a perturbation term $\Delta \varepsilon_r(\br)$:
\begin{align}\label{P60}
\varepsilon_r(\br) = \varepsilon_r^{(0)}(r) + \Delta \varepsilon_r(\br) ,
\end{align}
where
\begin{align}\label{P65}
\Delta \varepsilon_r(\br) = & \, - \left(n_1^2 - n_2^2 \right)\\[4pt]
& \times\left[ H \bigl( r- a  + a \,h(\hbr)  \bigr) - H\bigl(r- a  \bigr)\right] .
\end{align}
%
%
%
%
In the case of small deviations $\abs{h(\hbr)} \ll 1$ from the reference spherical surface, we can approximate \eqref{P65} with
\begin{align}\label{P70}
\Delta \varepsilon_r(\br) \cong & \, \left(n_1^2 - n_2^2 \right)\\[4pt]
& \times a  h(\hbr)\left[ \delta(r-a) - \frac{a h(\hbr)}{2}\, \delta'(r-a)\right],
\end{align}
where $\delta'(r-a) = \di \delta(r-a)/ \di r$ and we have expanded $\Delta \varepsilon_r(\br)$  to  second order because we plan to calculate quadratic corrections to the resonant wave numbers.
Evidently, there is a freedom in attributing the singular local terms in \eqref{P70} to either the internal $(r \leq a)$ or the external $(r>a)$ region \cite{PhysRevE.67.026215}. We choose to define $\Delta \varepsilon_r(\br)$ in the internal region  solely. This implies that we can define an effective potential $V(\epsilon, \br)$ as:
\begin{align}\label{PertV}
V(\epsilon, \br) =  \frac{\Delta \varepsilon_r(\br)}{n_1^2}
\equiv \epsilon \, V^{(1)}(\br) + \epsilon^2 \, V^{(2)}(\br),
\end{align}
where $\epsilon \geq 0$ is a formal  parameter serving to build a perturbation series with $V(0,\br)=0$, and  we have defined
\begin{equation}\label{PertVb}
\begin{split}
 V^{(1)}(\br) \equiv &  - v(k) \, a  \, h(\hbr)   \delta(r-a),  \\[4pt]
 V^{(2)}(\br) \equiv &  \; v(k) \, \frac{a^2  h^2(\hbr)}{2} \,  \delta'(r-a),
\end{split}
\end{equation}
with
\begin{equation}\label{PertVc}
v(k) \equiv k^2 (n_1^2 - n_2^2)/n_1^2.
\end{equation}

A caveat is in order here. The Debye potentials representation presented in Sec. \ref{IIA} is valid for electromagnetic fields in \emph{uniform} dielectric media. This condition is certainly satisfied by the physical dielectric bodies considered in this work. However, the use of the potential \eqref{PertV} introduces an effective inhomogeneity at $r=a$. As the Debye potentials representation is still valid inside the dielectric body $(r<a)$, in the spirit of perturbation theory it is reasonable to extend this representation to the whole region $r \leq a$, keeping in mind that this is an approximation.

\subsection{Kapur-Peierls perturbation theory}\label{KPpert}

According to the previous discussion, we consider now a perfect sphere whose refractive index is modified by a small perturbation  $V(\epsilon,\br)$ defined for
$r \leq a$ only. It must be put equal to $1$ at the end of the calculations.
Because of the both radial and angular dependence  of $V$ we have to generalize the radial equation \eqref{Compact} to
\begin{align}\label{Compact3D}
\bigl( \hD - k^2 \bigr) \Psi(\br) = 0,
\end{align}
where
\begin{align}\label{OpDb}
 \hD = & \; \frac{1}{n_1^2} \left( -\frac{\partial^2}{\partial r^2} + \frac{\hat{L}^2}{r^2} \right)+ V(\epsilon ,\br) \nonumber \\[4pt]
\equiv & \; \hD_0 + V(\epsilon ,\br).
\end{align}
As we deal with fields in the interior region only, in the remainder the index $j$ will be omitted. The Kapur-Peierls eigenvalue equation for the unperturbed operator $\hD_0$ reads as
\begin{align}\label{Aux2}
\bigl( \hD_0 - \lambda_{nl}(k) \bigr) \Phi_{nlm}(k,\br) & \; = 0,
\end{align}
where
\begin{align}\label{Numbers}
\Phi_{nlm}(k,\br) = \phi_{nl}(k,r) Y_{lm}(\theta,\phi),
\end{align}
and $\tilde{\Phi}_{nlm}(k,\br) = \tilde{\phi}_{nl}(k,r) Y_{lm}(\theta,\phi)$, with $n, \, l$ and $m$ being the so-called radial, azimuthal and magnetic numbers. The radial eigenfunctions $ \phi_{nl}(k,r)$ are defined as before by (\ref{Aux1}-\ref{KPfun1}). Since the
 boundary conditions \eqref{AuxBC} are independent of the magnetic number $m$, each eigenvalue  $\lambda_{nl}(k)$ is $2l+1$ times degenerate.

Now, suppose that $\epsilon \neq 0$. In this case when a wave with given radial, azimuthal and magnetic numbers $n,l$ and $m$ impinges upon the inhomogeneous dielectric sphere, it is scattered into many (possibly infinitely many) waves with different numbers $n',l'$ and $m'$. This occurs because the non spherically symmetric potential $V(\epsilon,\br)$ couples different modes of the field \cite{PhysRevA.93.042706}. Therefore, the ``single-channel'' Kapur-Peierls theory developed in the previous section is not directly applicable and the theory must be generalized (see, e.g., \cite{Kapur277,RevModPhys.31.893}). However, because of the spherically-symmetric surface of the inhomogeneous dielectric body, we still have well defined internal and external scattering regions characterized by $r \leq a$ and $r>a$, respectively. In this case it is not difficult to show \cite{PhysRevA.67.013805} that the original Kapur-Peierls equation \eqref{Aux1} can be replaced by
the new eigenvalue equation
\begin{align}\label{Aux3}
\bigl( \hD - \Lambda_{nlm}(k) \bigr) \Psi_{nlm}(k,\br) & \; = 0,
\end{align}
and the fixed-point equation \eqref{Join5} becomes
\begin{align}\label{Res10}
K_{nlm}^2 = \Lambda_{n l m}(K_{nlm}),
\end{align}
which must reduce to $k_{nl}^2 = \lambda_{n l}(k_{nl})$ for $\epsilon =0$. However, it is important to keep in mind that while \eqref{Join5} is an \emph{exact} relation, equation \eqref{Res10} rests upon the approximation of replacing a near-spherical homogeneous dielectric body with an inhomogeneous spherical one.

Now, according to Rayleigh-Schr\"{o}dinger  perturbation theory suitably adapted to the case of a biorthogonal basis \cite{PhysRevC.6.114,Brody2014}, we assume that $\Psi_{nlm}(k,\br)$ and $\Lambda_{nlm}(k)$  can be expanded in powers of $\epsilon$:
\begin{align}\label{p130A}
\Psi_{nlm}(k,\br)= & \;  \Psi_{nlm}^{(0)}(k,\br)  \nonumber \\[4pt]
& +  \epsilon \, \Psi_{nlm}^{(1)}(k,\br)  + \epsilon^2 \Psi_{nlm}^{(2)} (k,\br) + \ldots \, ,
\end{align}
\begin{align}\label{p130B}
\! \! \!  \Lambda_{n l m} (k) =   \Lambda_{n l m}^{(0)}(k)    +  \epsilon \, \Lambda_{n l m}^{(1)}(k)  +  \epsilon^2 \Lambda_{n l m}^{(2)}(k) + \ldots ,
\end{align}
where $\Lambda_{n l m}^{(0)}(k) = \lambda_{n l}(k)$. 
Similarly, we write
\begin{align}\label{Res20}
K_{nlm} = K_{nlm}^{(0)} + \epsilon \, K_{nlm}^{(1)} + \epsilon^2  K_{nlm}^{(2)} + \ldots,
\end{align}
with $K_{nlm}^{(0)} = k_{nl}$.
Suppose that by using standard techniques we have calculated the first two terms of the expansion \eqref{p130B}.
 Substituting \eqref{Res20} into \eqref{Res10} and using \eqref{p130B}, we obtain
\begin{align}\label{Res30}
 \bigl(  k_{nl} &  + \epsilon \, K_{nlm}^{(1)} + \epsilon^2  K_{nlm}^{(2)} + \ldots \bigr)^2 \nonumber \\[5pt]
  = & \; \lambda_{n l}\bigl( k_{nl} + \epsilon \, K_{nlm}^{(1)} + \epsilon^2  K_{nlm}^{(2)} + \ldots \bigr) \nonumber \\[5pt]
& +  \epsilon \, \Lambda_{n l m}^{(1)} \bigl( k_{nl} + \epsilon \, K_{nlm}^{(1)} + \epsilon^2  K_{nlm}^{(2)} + \ldots \bigr)   \nonumber \\[5pt]
& +  \epsilon^2 \Lambda_{n l m}^{(2)} \bigl( k_{nl} + \epsilon \, K_{nlm}^{(1)} + \epsilon^2  K_{nlm}^{(2)} + \ldots \bigr)  + \ldots
\end{align}
Expanding the functions on the right side of this equation in Taylor series around $\epsilon =0$ and equating the terms with the same powers of $\epsilon$  on both sides we find, up to and including second-order terms,
\begin{subequations}\label{p170}
\begin{align}
k_{nl}^2 =  & \;\lambda_{n l}(k_{nl}),  \label{p170A} \\[4pt]
K^{(1)}_{n l m} =  & \; \frac{ \Lambda_{n l m}^{(1)}(k_{nl})}{\displaystyle{2 k_{nl} - \left. \frac{\di \lambda_{n l}(k)}{\di k}\right|_{k=k_{nl}}}}  , \label{p170B} \\[4pt]
K^{(2)}_{n l m} =  & \; \frac{ 1}{\displaystyle{2 k_{nl} - \left. \frac{\di \lambda_{n l}(k)}{\di k}\right|_{k=k_{nl}}}}
\Biggr. \nonumber \\[4pt]
& \; \times \Biggl\{ \Lambda_{n l m}^{(2)}(k_{nl}) + K^{(1)}_{n l m} \left. \frac{\di \Lambda_{n l m}^{(1)}(k)}{\di k}\right|_{k=k_{nl}} \nonumber \\[4pt]
& \;  \Biggl.  - \left({K^{(1)}_{n l m}}\right)^2 \left[ 1 - \frac{1}{2}\left. \frac{\di^2 \lambda_{n l}(k)}{\di k^2}\right|_{k=k_{nl}} \right]
\Biggr\}.
\label{p170C}
\end{align}
\end{subequations}
The two terms
\begin{align}\label{Res40}
\left.\frac{\di \lambda_{n l}(k)}{\di k}\right|_{k=k_{nl}} \qquad \text{and} \qquad \frac{1}{2} \! \! \left.\frac{\di^2 \lambda_{n l}(k)}{\di k^2}\right|_{k=k_{nl}},
\end{align}
can be calculated substituting the Taylor expansion of $\lambda_{nl}(k)$ around $k=k_{nl}$,
%
%
%
%
%
into $F_l \bigl( n_1 \sqrt{\lambda(k)} \, a, k a \bigr) =0$,
%
%
%
%
and equating to zero the terms with the same power of $\left( k - k_{nl} \right) $. After a straightforward calculation we find:
%
%
\begin{align}
\left.\frac{\di \lambda_{n l}(k)}{\di k}\right|_{k=k_{nl}} = & \; - \frac{2 k_{nl}}{n_1} \, \rho(k_{nl}), \label{Res50A} 
\end{align}
and
\begin{align}
\frac{1}{2} \! \! \left.\frac{\di^2 \lambda_{n l}(k)}{\di k^2}\right|_{k=k_{nl}} \!  = & \; \frac{\rho^2(k_{nl})}{n_1^2} -\frac{ k_{nl} a}{n_1} \frac{1}{\displaystyle{\left. \frac{\partial F_l(z,  k_{nl} a)}{ \partial z} \right|_{z = n_1 k_{nl} a}}} \nonumber \\[4pt]
& \times \Biggl[ \frac{\partial^2 F_l(z,  w)}{ \partial w^2} - 2  \frac{\partial^2 F_l(z,  w)}{\partial z \, \partial w} \, \rho(k_{nl})
\nonumber \\[4pt]
&  +\frac{\partial^2 F_l(z,  w)}{ \partial z^2} \, \rho^2(k_{nl}) \Biggr]\!\!{\phantom{\Bigl|}}_{\substack{z = n_1 k_{nl} a \\\!\!\! \! \!\! \!\!  w = k_{nl} a}}  ,
 \label{Res50B}
\end{align}
%
%
where
\begin{align}\label{Res51}
\rho(k_{nl}) \equiv \frac{\displaystyle{\left. \frac{\partial F_l(n_1  k_{nl} a,w)}{ \partial w} \right|_{w = k_{nl} a} }}{\displaystyle{\left. \frac{\partial F_l(z,  k_{nl} a)}{ \partial z} \right|_{z = n_1 k_{nl} a}}} .
\end{align}
Incidentally, we note that iterating this procedure it is possible to calculate the function $\lambda_{n l}(k)$ in the neighborhood of any point $k_{nl}$  with the desired degree of accuracy.

Equations \eqref{p170} are the main result of this section; they formally solve completely our problem. The zeroth-order equation \eqref{p170A} simply reproduces the resonances of the unperturbed system. The other two equations gives first- and second-order corrections in terms of the two functions $\Lambda_{n l m}^{(1)}(k)$ and $\Lambda_{n l m}^{(2)}(k)$ that will be explicitly calculated in the next section. The physical meaning of the denominator in \eqref{p170B} is explained in \cite{PhysRevA.7.1288}; it amounts to a renormalization factor connecting Kapur-Peierls eigenmodes with Gamow (i.e., decaying) modes. The second and third term within the curly brackets in \eqref{p170C}  represent  second-order corrections that, in general, should not be neglected with respect to $\Lambda_{n l m}^{(2)}(k_{nl})$. 

\section{Rayleigh-Schr\"{o}dinger perturbation theory}\label{PerTheory}


In this section we use Rayleigh-Schr\"{o}dinger perturbation theory to find the optical resonances of a deformed dielectric sphere. The only (trivial) difference with respect to familiar quantum perturbation theory is the use of biorthogonal bases \cite{PhysRevC.6.114,Brody2014}.

 {Let us consider  a specific unperturbed resonant wavenumber $k_{nl}$ where $n$ and $l$ have now \textbf{fixed} values. The corresponding unperturbed Kapur-Peierls eigenvalue is $\lambda_{nl}(k)$, which we assume to be non-degenerate at the interesting values of $k$. Here, with ``non-degenerate'' we mean that there is single \emph{radial}  wavefunction $\phi_{nl}(k,r)$ defined by \eqref{Numbers} and associated with the eigenvalue $\lambda_{nl}(k)$ via the eigenproblem \eqref{Aux2}  \cite{PhysRevA.7.1288}.  However, there are $2l+1$ different solutions  of \eqref{Aux2} associated with the same eigenvalue  $\lambda_{nl}(k)$, which are obtained by multiplying the unique radial wavefunction $\phi_{nl}(k,r)$ by the $2l+1$ \emph{angular-dependent} spherical harmonics $ Y_{lm}(\hbr)$:
\begin{multline}\label{Deg1}
\{ \Phi_{nlm}(k,\br)\} \\ =  \{\phi_{nl}(k,r) Y_{l,-l}(\hbr), \ldots, \phi_{nl}(k,r) Y_{ll}(\hbr) \}.
\end{multline}
}
These solutions span a $(2l+1)$-dimensional degenerate subspace, which we call  $\mathcal{D}_{nl}$. According to degenerate perturbation theory, we build the new set of eigenfunctions  $\{ \Phi_{nlm}^{\mathcal{D}}(k,\br)\} \in \mathcal{D}_{nl}$, defined by
\begin{multline}\label{Deg2}
\{ \Phi_{nlm}^{\mathcal{D}}(k,\br)\} \\ =  \{\phi_{nl}(k,r) \mathcal{Y}_{l,-l}(\hbr), \ldots, \phi_{nl}(k,r) \mathcal{Y}_{ll}(\hbr) \},
\end{multline}
where
\begin{align}\label{Deg3}
\mathcal{Y}_{lm}(\hbr) \equiv \sum_{m' =-l}^l C_{lm}^{m'} Y_{lm'}(\hbr).
\end{align}
As usual, the coefficients $C_{lm}^{m'}$ can be determined solving the eigenvalue equation
\begin{multline}\label{Deg4}
 \sum_{m'' =-l}^l \Bigl(\tilde{\Phi}_{nlm'}, V^{(1)} (\br)\, \Phi_{nlm''} \Bigr)_\br C_{lm}^{m''} \\ = \Lambda^{(1)}_{nlm}(k) \, C_{lm}^{m'},
\end{multline}
where here and hereafter we use the shorthand notation
%
%
%
%
%
\begin{equation}\label{Deg5}
\begin{split}
\bigl( u, w \bigr)_{\br} \equiv &   \int\limits_0^a  \di r   \int\limits_0^{2 \pi} \di \phi  \int\limits_0^{\pi}\di \theta  \sin \theta \, u^*(r, \theta , \phi) w(r, \theta , \phi), \\[6pt]
\bigl( u, w \bigr)_{\hbr} \equiv & \int\limits_0^{2 \pi} \di \phi  \int\limits_0^{\pi}\di \theta  \sin \theta \,  u^*( \theta , \phi) w( \theta , \phi),
\end{split}
\end{equation}
%
%
(note that the radial differential is  $\di r$ and not $r^2 \di r$.)
Substituting \eqref{Numbers} and \eqref{PertVb} into \eqref{Deg4} and solving it for $\Lambda^{(1)}_{nlm}(k)$, we obtain the first-order correction to $k_{nl}$:
\begin{align}\label{Deg6}
\Lambda^{(1)}_{nlm}(k) = - a \, v(k) \, \phi_{nl}^2(k,a) \,  \ell_{lm},
\end{align}
%
%
where $v(k) = k^2 (n_1^2 - n_2^2)/n_1^2 $ and
\begin{align}\label{Deg7}
\ell_{lm} \equiv \bigl( \mathcal{Y}_{lm} , h(\hbr) \mathcal{Y}_{lm} \bigr)_\hbr,
\end{align}
with $m= -l, -l+1, \ldots,l$. This result allows us to find the first-order corrections $K^{(1)}_{nlm}$ by substituting \eqref{Deg6}, evaluated at $k = k_{nl}$, into \eqref{p170B}.

It should be noted that although  $\ell_{lm} $ is real by definition, $\Lambda_{\nu l m}^{(1)}(k)$ may be not, because $\phi_{n l}^2(k,a)$ is, in general, a complex number. However, using \eqref{Poles} and \eqref{KPfun1}  it is not difficult to show that for TE polarization,
\begin{align}\label{Deg7c}
\frac{- a \, v(k_{nl}) \, \phi_{nl}^2(k_{nl},a)}{\displaystyle{2 k_{nl} - \left. \frac{\di \lambda_{n l}(k)}{\di k}\right|_{k=k_{nl}}}} = - k_{nl},
\end{align}
and \eqref{p170B} becomes
\begin{align}\label{Deg7b}
K^{(1)}_{nlm}(k_{nl}) = - k_{nl} \, \ell_{lm}.
\end{align}
Since $\ell_{lm}$ is a real number, from \eqref{Deg7b} and \eqref{Q} it follows that the $Q$ factor of TE waves is not affected by first order corrections. However, for TM polarization  a simple expression as \eqref{Deg7b} does not exist because the left side of \eqref{Deg7c} displays a complicated functional dependence on $k_{nl}$ that will not be reported here. This implies that the $Q$ factor of TM waves may be affected  by first-order corrections.

The independence of $\ell_{lm}$ from the wave number $k$, the polarization $p$, the refractive index $n_1$ and from the radial part of the radial function $\phi_{nl}(k,r)$, is a surprising result of first-order perturbation theory, which was discovered already in the nineties of last century \cite{PhysRevA.41.5187,PhysRevA.41.5199}.

\subsection{Discussion of the first-order corrections}\label{PerTheory2}

From the definition \eqref{Deg7} and  \eqref{ThermalH} it follows that
$\ell_{lm}$ is a real number independent of $k$ and coincides with the $m$-th eigenvalue of the $(2l+1) \times (2l+1)$ Hermitean matrix   $H_l$ defined by
\begin{align}\label{Deg7g}
 [H_l]_{m m'} = ({Y}_{lm} , h(\hbr) {Y}_{lm'})_\hbr, \quad (m,m' = -l,  \ldots, l).
\end{align}
Moreover, for fixed $l$ and $m$ the coefficients  $C_{lm}^{m'}$ in \eqref{Deg3} coincide with the components  $(C_{lm}^{-l}, C_{lm}^{-l+1}, \ldots, C_{lm}^{l})$ of the  $m$-th  eigenvector $\mathbf{C}_{lm}$ associated with $\ell_{lm}$, namely $H_l \mathbf{C}_{lm} = \ell_{lm} \mathbf{C}_{lm}$.

The matrix elements \eqref{Deg7g} can be calculated from \eqref{ThermalH} and expressed in terms of the Wigner $3j$-symbols \cite{Wigner3j-Symbol} as:
\begin{widetext}
\begin{align}\label{Deg8}
({Y}_{lm} ,  h(\hbr) {Y}_{lm'})_\hbr = & \;   (-1)^{m}(2 l +1)\sum_{L=2}^\infty
\sqrt{\frac{2L+1}{4 \pi}}
\begin{pmatrix}
                            l & l & L \\
                            0 & 0 & 0 \\
                          \end{pmatrix} \sum_{M=-L}^L h_{LM}
\begin{pmatrix}
                            l   & l   & L \\
                            -m & m' & M \\
                          \end{pmatrix} \nonumber \\[4pt]
%
%
= & \;   (-1)^{m}(2 l +1)\sum_{{l'}=1}^{l}h_{2{l'},m-m'}
\sqrt{\frac{4{l'}+1}{4 \pi}}
\begin{pmatrix}
                            l & l & 2{l'} \\
                            0 & 0 & 0 \\
                          \end{pmatrix}
\begin{pmatrix}
                            l   & l   & 2{l'} \\
                            -m & m' & m-m' \\
                          \end{pmatrix},
\end{align}
\end{widetext}
where the second expression follows from
 the properties of the $3j$-symbols requiring that  only terms with $L$  even, $L \leq 2 l$ and $M = m - m'$, contribute to $[H_l]_{m m'}$. This means that at first-order level the resonance $k_{nl}$ is not affected by ``rapid'' surface fluctuations characterized by $L >2 l$.

The matrix $H_l$ can be huge. For a $\Heq$ droplet of radius $a = 1 \, \text{mm}$, refractive index $n_1 \approx 1.03$ and illuminated by light of wavelength $\lambda = 1 \, \mu\text{m}$ in vacuum, the value of $l$ is around $l \approx 2 \pi a n_1/\lambda \approx 6500$ \cite{Oraevsky}. Diagonalizing a matrix of dimension $\sim 10^4 \times 10^4$ with sufficient accuracy may be a serious task depending on the distribution of the matrix elements and on available computational resources. 
We discuss a way to  circumvent these problems in appendix \ref{Hldiag}.

\subsection{Second-order corrections}\label{PerTheory3}

Because of the form \eqref{PertV} of the perturbation, the second-order correction $\Lambda^{(2)}_{nlm}(k)$ contains two terms:
\begin{widetext}
\begin{align}\label{Second10}
\Lambda^{(2)}_{nlm}(k) = & \;  \Bigl( \tilde{\Phi}_{nlm}^\mathcal{D} , V^{(2)}(\br )   \Phi_{nlm}^\mathcal{D} \Bigr)_\br
 + \sum_{n',l',m' \notin \mathcal{D}_{nl}} \frac{\Bigl( \tilde{\Phi}_{nlm}^\mathcal{D} , V^{(1)}(\br )   \Phi_{n'l'm'} \Bigr)_\br\Bigl( \tilde{\Phi}_{n'l'm'} , V^{(1)}(\br )   \Phi_{nlm}^\mathcal{D} \Bigr)_\br}{\lambda_{nl}(k) - \lambda_{n' l'}(k)} \nonumber \\[4pt]
\equiv & \; A + B.
\end{align}
\end{widetext}
Using \eqref{PertVb} and \eqref{Deg2} we can rewrite the first term in the equation above as:
\begin{align}\label{Second20}
A =  - a^2 v(k)\phi_{nl}(k,a) \phi_{nl}'(k,a) \, T_{lm},
\end{align}
where we have defined
\begin{align}\label{Second25}
T_{lm} \equiv   \left(\mathcal{Y}_{lm}, h^2(\hbr)\mathcal{Y}_{lm} \right)_\hbr,
\end{align}
 and $\phi_{nl}'(k,a) = \left. \di \phi_{nl}(k,r)/\di r \right|_{r=a} $.
%
Similarly, after a straightforward calculation we obtain for the second term,
\begin{align}\label{Second30}
B = & \; a^2 v^2(k)\phi_{nl}^2(k,a)  \nonumber \\[4pt]
& \times {\sum_{l'}}' \Biggl[ {\sum_{n'}}' \frac{\phi_{n'l'}^2(k,a)}{\lambda_{nl}(k) - \lambda_{n' l'}(k)} \, T_{lm}^{\,l'} \Biggr] ,
\end{align}
where
\begin{align}\label{Second30bis}
T_{lm}^{\,l'}  \equiv & \;  \sum_{m' = -l'}^{l'} \abs{\left(Y_{l'm'}, h(\hbr)\mathcal{Y}_{lm} \right)_\hbr}^2,
\end{align}
and the prime symbols above the sums in $l'$ and $n'$ dictate the exclusion of the term with $(n',l') = (n,l)$. These sums are really formidable and, for high values of $l$, represent  a hard numerical challenge. However, a huge simplification can be made by noticing that after replacing everywhere $k$ with $k_{nl}$,
 the sum with respect to $n'$ with $l'\neq l$ in \eqref{Second30} can be rewritten as
\begin{align}\label{Second40}
\sum_{n'} \frac{\phi_{n'l'}^2(k_{nl},a)}{k_{nl}^2 - \lambda_{n' l'}(k_{nl})} = G_{l'}(k_{nl},a,a) ,
\end{align}
where \eqref{p170A} has been used and $G_{l'}(k_{nl},a,a) \equiv G_{l'}(k_{nl})$ is the Green function defined in sec. \ref{KPa3}. Comparing this equation with \eqref{Join3} and \eqref{Join4} we  obtain a closed expression for the infinite sum  \eqref{Second40}:
\begin{align}\label{Second50}
G_{l'}(k_{nl}) =   \frac{a \, n_1^2 \, j_{l'}(n_1 k_{nl}a) \, h_{l'}^{(1)}(k_{nl}a)}{f_{l'}(k_{nl}a)},
\end{align}
where $f_{l'}(k_{nl}a)$ is the Jost function defined by \eqref{Jost}. Therefore, we can eventually rewrite \eqref{Second30} as:
\begin{align}\label{Second30ter}
B = & \; a^2 v^2(k)\phi_{nl}^2(k,a) \Biggl[ \sum_{n'\neq n} \frac{\phi_{n'l}^2(k_{nl},a)}{k_{nl}^2 - \lambda_{n' l}(k_{nl})} \, T_{lm}^{\,l} \nonumber \\[4pt]
& + \sum_{l'\neq l} G_{l'}(k_{nl}) \, T_{lm}^{\,l'} \Biggr] .
\end{align}
Eventually, the awkward double sum in \eqref{Second30} was split in two simpler single sums, one with respect to $n' \neq n$ and the other with respect to $l' \neq l$.

\subsection{Summary of the main results and discussion}\label{PerTheory4}

Collecting the results above we can summarize our main findings as follows. The first-order correction to the unperturbed resonant wavenumber $k_{nl}$ is given by \eqref{Deg6} evaluated at $k = k_{nl}$, that is
\begin{align}\label{Deg6b}
\Lambda^{(1)}_{nlm}(k_{nl}) = - a \, v(k_{nl}) \, \phi_{nl}^2(k_{nl},a) \,  \ell_{lm},
\end{align}
where
\begin{align}\label{Deg7f}
\ell_{lm} \equiv \bigl( \mathcal{Y}_{lm} , h(\hbr) \mathcal{Y}_{lm} \bigr)_\hbr,
\end{align}
is independent of $k_{nl}$ and
\begin{itemize}
  \item $v(k_{nl}) = k_{nl}^2 (n_1^2 - n_2^2)/n_1^2 $,
  \item $\phi_{nl}(k_{nl},a) = \frac{{\sqrt{\frac{2}{a}}\, j_l \left( n_1  k_{nl}  a \right)}}{\sqrt{j_{l}^{\,2}( n_1 k_{nl} a) - j_{l-1}( n_1 k_{nl}  a) j_{l+1}( n_1 k_{nl}  a) }}  $,
  \item $\displaystyle{h(\hbr) =  \sum_{L=2}^\infty \sum_{M=-L}^L h_{LM} Y_{LM}(\hbr)}$,
  \item $\displaystyle{\mathcal{Y}_{lm}(\hbr) \equiv \sum_{m' =-l}^l C_{lm}^{m'} Y_{lm'}(\hbr)}$ ,
\end{itemize}
where (\ref{KPfun1},\ref{ThermalH},\ref{PertVc}) and \eqref{Deg3} have been used.
 {The first-order correction \eqref{Deg6b} depends on the index $m$ via  the term $\ell_{lm}$. This may or may not fully remove the degeneration of the unperturbed states \eqref{Deg2} according to the form of the deformation $h(\hbr)$. For example, we shall see later that when the sphere is deformed into an ellipsoid of revolution, only the states that differ by the sign of $m$ remain degenerate. Therefore, in this case we pass from a $2 l + 1$ degenerate subspace (which, as previously discussed, can be huge) to a smaller $2$-dimensional space}.

 The second-order correction is obtained from (\ref{Second10}-\ref{Second30ter}) and it is equal to
\begin{align}\label{Second60}
 \Lambda^{(2)}_{nlm}(k_{nl})\! = &  -a^2 \, v^2(k_{nl}) \phi_{nl}^2(k_{nl}a)  \nonumber \\[4pt]
& \times \Biggl[ \frac{p\, c_{\,l}(k_{nl}a)}{v(k_{nl})} \, T_{lm} - \!\! \sum_{n'\neq n} \frac{\phi_{n'l}^2(k_{nl},a)}{k_{nl}^2 - \lambda_{n' l}(k_{nl})} \, T_{lm}^{\,l}  \nonumber \\[4pt]
&- \sum_{l' \neq l} G_{l'}(k_{nl}) \, T_{lm}^{\,l'} \Biggr],
\end{align}
where $p =1$ for TE polarization, $p = n_1^2/n_2^2$ for  TM polarization and   (\ref{Cl}, \ref{AuxBC}) have been used. The terms $c_{\,l}(k_{nl}a)$, $T_{lm}$ and $T_{lm}^{\,l'}$ are defined by (\ref{Cl},\ref{Second25}) and \eqref{Second30bis}, respectively. The three coefficients $\ell_{lm}, \, T_{lm}$ and $T_{lm}^{\,l'} $, contain \textbf{all} the information about the perturbation. They are the fundamental quantities that must be calculated for a given perturbation function $h(\hbr)$ and all of them are independent from the resonant wave number $k_{nl}$, the polarization $p$, the refractive index $n_1$ and the radial part of the radial function $\phi_{nl}(k_{nl},r)$. These coefficients, which are independent of the radial wave function, depend upon the eigenvectors and the eigenvalues of the matrix $H_l$, which is defined by \eqref{Deg7g}, via the functions $\mathcal{Y}_{lm}(\hbr)$ given by \eqref{Deg3}. The meaning of the various sums in \eqref{Second60} is graphically illustrated in Fig. \ref{figS} below.
%
%
\begin{figure}[!ht]
\centerline{\includegraphics[scale=3,clip=false,width=1\columnwidth,trim = 0 0 0 0]{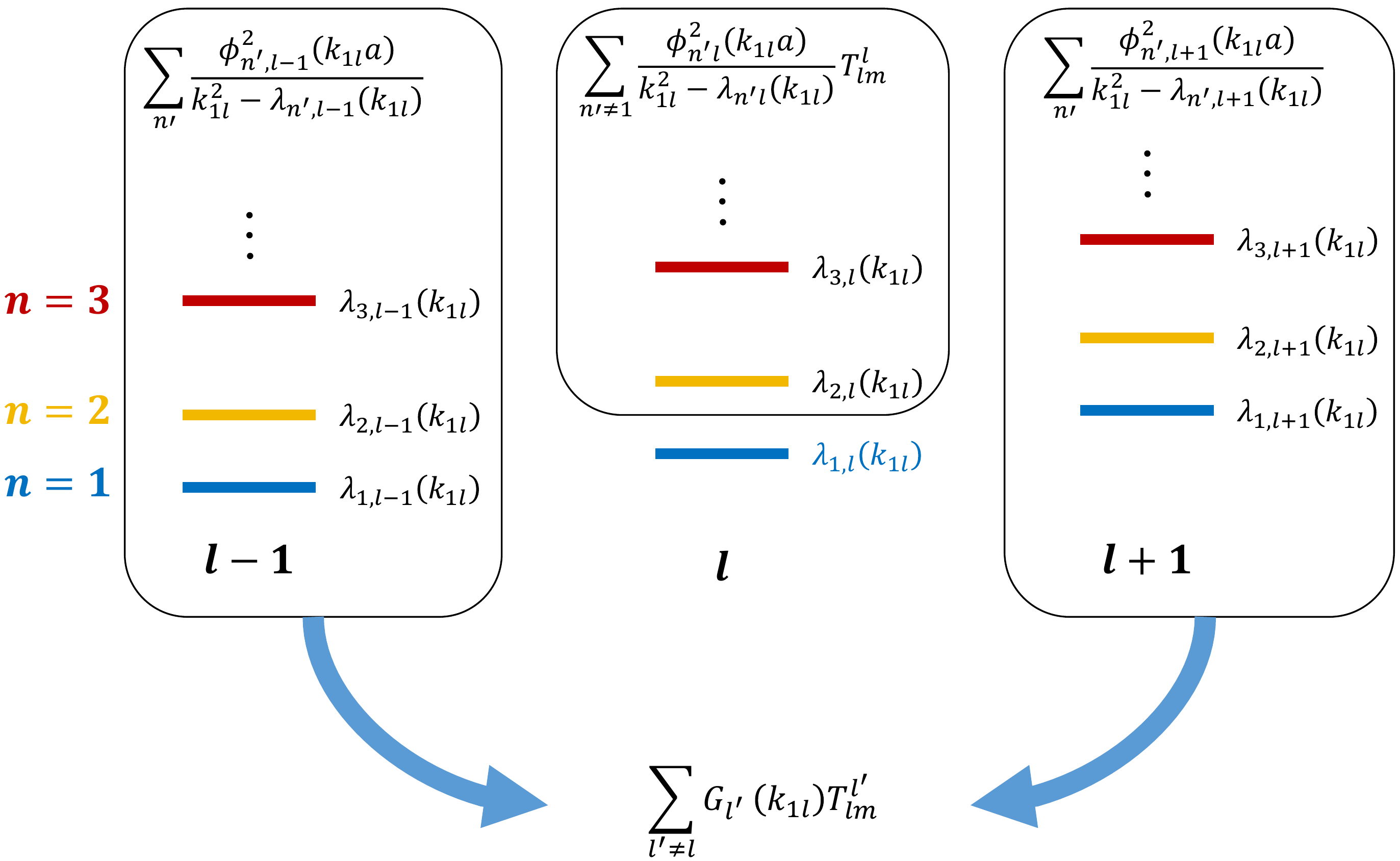}}
\caption{\label{figS}  Qualitative representation of the Kapur-Peierls spectrum. As an example, the KP eigenvalues $\lambda_{n'l'}(k_{nl})$ with azimuthal number $l' = l-1, \, l,l+1$ are depicted as horizontal bands colored  blue ($n'=1$), orange ($n'=2$) and red ($n'=3$) according to the value of $n'$.
$k_{1l} = \sqrt{\lambda_{1l}(k_{1l})}$ is the unperturbed KP eigenvalue whose second-order corrections caused by the sphere deformation we want to calculate using  \eqref{Second60}. Each black frame around a portion of the spectrum encloses all the eigenvalues contributing to the sum inside the frame. Finally, the sums with $l' \neq l$ are again summed as indicated by the thick blue arrows. Clearly, the biggest contribution to each sum comes from those KP eigenvalues such that $\abs{k_{1l}^2 - \lambda_{n'l'}(k_{1l})}\ll 1$.}
\end{figure}
%

Once $\Lambda^{(1)}_{nlm}(k_{nl})$ and $\Lambda^{(2)}_{nlm}(k_{nl})$ have been calculated, they must be substituted into \eqref{p170} to obtain the perturbed characteristic value $K_{nlm}$ correct up to second-order terms. From $K_{nlm}$ we can  calculate the central frequency $\omega_{nlm}$ and the bandwidth $\Delta\omega_{nlm}$ of the resonance identified by $n,l,m$:
\begin{align}\label{Second65}
\omega_{nlm} = c  \re K_{nlm}, \qquad \Delta \omega_{nlm} = -2 c  \im K_{nlm},
\end{align}
and  the $Q$ factor as well using the definition \eqref{Q}. The result, up to and including second-order terms (here we put $\epsilon=1$), is:
\begin{align}\label{Second70}
\frac{1}{Q(K_{nlm})} \cong &  \;   \frac{1}{Q(k_{nl})} \Biggl[ 1 + R^{(1)}_{nlm}  \nonumber \\[4pt]
& + \left(  R^{(2)}_{nlm} - \frac{\re K^{(1)}_{nlm}}{\re k_{nl}} \, R^{(1)}_{nlm} \right) \Biggr],
\end{align}
where we have defined
\begin{align}\label{Second80}
R^{(\beta)}_{nlm} \equiv  \frac{\im K^{(\beta)}_{nlm}}{\im k_{nl}} - \frac{\re K^{(\beta)}_{nlm}}{\re k_{nl}}, \qquad (\beta=1,2).
\end{align}
From \eqref{Deg7b} it follows that for TE waves $R^{(1)}_{nlm} = 0$ and  \eqref{Second70} reduces to the very simple form
\begin{align}\label{Second70b}
\frac{1}{Q(K_{nlm})} \cong   \frac{1}{Q(k_{nl})} \left( 1 + R^{(2)}_{nlm} \right).
\end{align}
%

 %

\subsubsection{Examples}\label{Examples}

Our results are in agreement with previous works where first-order perturbation theory for leaking electromagnetic modes in open systems was developed \cite{PhysRevA.41.5187,PhysRevA.41.5199}. To see this, let us consider the following two examples.

\subsubsection{Equatorial bulge}

Consider a TE excitation of the droplet, this sets $p=1$. Suppose that $h(\hbr)$ describes an ellipsoid of revolution  with polar and equatorial radii $a_P$ and $a_E> a_P$, respectively, with $a_P \, a_E^2=a^3$. The ellipticity (or, eccentricity) of this ellipsoid is denoted $e$ and defined by
\begin{align}\label{Ellipticity}
e = \sqrt{1-\frac{a_P^2}{a_E^2}} \, .
\end{align}
The surface profile function of the ellipsoid of revolution is
\begin{align}\label{SurProfObl}
a + a \, h(\hbr) =  \frac{a_P \, a_E}{\sqrt{a_E^2 \cos^2 \theta + a_P^2 \sin^2 \theta}} ,
\end{align}
which, when $e \ll 1$, can be approximated by
\begin{align}\label{SurProfObl2}
 h(\hbr) \cong - \frac{e^2}{12}\bigl[1+3 \cos (2 \theta) \bigr]=-\frac{2}{3}\sqrt{\frac{\pi}{5}} \, e^2 Y_{20}(\hbr).
\end{align}
Then, from this equation and  \eqref{Deg8} it follows that
\begin{align}\label{Deg9}
({Y}_{lm} ,  h(\hbr) {Y}_{lm'})_\hbr =   \delta_{m m'} \,\frac{e^2}{3} \frac{l(l+1)-3 m^2}{4 l(l+1)-3} .
\end{align}
Substituting \eqref{Deg9} into \eqref{Deg7b} we obtain, for $l \gg 1$,
\begin{align}\label{Deg10}
\frac{K^{(1)}_{nlm}(k_{nl})}{k_{nl}} = - \frac{e^2}{12} \left[ 1-\frac{3 m^2}{l(l+1)}\right],
\end{align}
which is in perfect agreement with \cite{PhysRevA.41.5187} (note: because of a different definition, the parameter $e$ used in \cite{PhysRevA.41.5187} is equal to our $e^2/2$).

\subsubsection{Shrinking sphere}

As a second example, consider as a perturbation the change of the radius of the sphere from $a$ to $b<a$, such that $a-b \equiv \delta a \ll a$. Let $z_{nl}$ be a root of the equation \eqref{Poles} $f_l(z)=0$ with $p=1$ (TE polarization) and denote with $k_{nl}(a) \equiv z_{nl}/a$ and $k_{nl}(b) \equiv z_{nl}/b$ the two corresponding resonances of the bigger and smaller cavity. Then, trivially,
\begin{align}\label{Deg20}
k_{nl}(b) = & \; \frac{k_{nl}(a)}{\displaystyle{1-\frac{\delta a}{a}}} \nonumber \\[4pt]
\cong & \; k_{nl}(a) \left( 1 + \frac{\delta a}{a} + \frac{\delta a^2}{a^2} + \ldots \right)\nonumber \\[4pt]
\equiv & \; k_{nl}(a) + K^{(1)}_{nlm} + K^{(2)}_{nlm} + \ldots
\end{align}
The surface profile function \eqref{cavEq} of the sphere of radius $b$ is evidently $g(\hbr) =b$. This implies that $h(\hbr) = - \delta a/a$. The matrix $H_l$ has elements $[H_l]_{m m'} = -(\delta a/a) \delta_{m m'}$ and, therefore, $\ell_{lm} = -\delta a/a$. A straightforward calculation shows that
\begin{align}\label{Deg30}
T_{lm} = (\delta a/a)^2 \quad \text{and} \quad T_{lm}^{l'} = (\delta a/a)^2 \, \delta_{l l'}.
\end{align}
Then, \eqref{Deg7} yields
\begin{align}\label{Deg40}
K^{(1)}_{nlm} = k_{nl}(a) \, \frac{\delta a}{a},
\end{align}
in perfect agreement with \eqref{Deg20}. From \eqref{p170}  and \eqref{Second60} we obtain:
\begin{align}\label{Deg50}
k_{nl}(b) \cong k_{nl}(a) \left( 1 +   \frac{\delta a}{a} + N_{nl} \frac{\delta a^2}{a^2}  \right) ,
\end{align}
where $N_{nl}$ is a finite complex-valued numerical coefficient that can be calculated explicitly once $n$ and $l$ have been fixed. Equation \eqref{Deg50} is in agrement with   \eqref{Deg20} up to $O({\delta a^2}/{a^2})$ corrections.

\section{Summary}\label{Summary}

In this work we have used the Kapur-Peierls formalism, originally developed in the context of nuclear scattering theory, to find the optical resonances of almost spherical dielectric objects, such as liquid drop. This permitted us to develop a second-order perturbation theory for the electromagnetic Debye potentials describing the light fields. We have thus found analytical formulas for the complex characteristic values of the resonances, whose real and imaginary parts are proportional to, respectively, the central frequency and the bandwidth of the optical resonance. When limited to first-order perturbation theory, our results are in perfect agreement with older results  \cite{PhysRevA.41.5187}. The present work provides the basis for applications
of  our technique to various optical and optomechanical systems.

\begin{acknowledgments}

This work was supported by the European Union’s Horizon 2020 research and innovation programme under grant agreement No 732894 (FET Proactive HOT), (A.A. and F.M.)
A.A. is grateful to Carlos Viviescas for useful discussions. J.H. acknowledges Charles Brown for his contribution and
support from the W. M. Keck Foundation Grant No. DT121914, AFOSR Grant No. FA9550-15-1-0270, DARPA Grant No. W911NF-14-1-0354, and NSF Grant No. 1205861. This project was made possible through the support of a grant from the John Templeton Foundation.
\end{acknowledgments}

\appendix

\section{Spherical Bessel and Hankel functions}\label{Bessel}

Spherical Bessel and Hankel functions are frequently encountered in scattering theory. Many of their properties can be found in Appendix A of \cite{Grandy} and in Appendix A.9 of \cite{GalindoI}.
Some properties utilized in this paper are:
\begin{enumerate}
  \item Recurrence:
\begin{align}\label{recur1}
j_{l-1}(z) = & \;  \frac{2 l +1}{z} j_l(z) - j_{l+1}(z).
\end{align}
  \item Differentiation:
\begin{align}\label{diff}
z \, j_{l+1}(z)   =  (l+1) j_l(z) - \frac{d }{d z} \left[ z j_l(z)\right],
\end{align}
  \item Parity:
\begin{subequations}\label{parity}
\begin{align}
j_l(-z) = & \; (-1)^l j_l(z), \label{parityA} \\[4pt]
y_l(-z) = & \; (-1)^{l+1} y_l(z), \label{parityB} \\[4pt]
h^{(1)}_l(-z) = & \; (-1)^l h^{(2)}_l(z). \label{parityC}
\end{align}
\end{subequations}
  \item Analytic continuation:
\begin{subequations}\label{AnCont}
\begin{align}
\bigl[ j_l(z) \bigr]^* = & \; j_l(z^*), \label{AnContA} \\[4pt]
\bigl[ y_l(z) \bigr]^* = & \; y_l(z^*), \label{AnContB} \\[4pt]
 j_l(-z^*) = & \; (-1)^l \bigl[ j_l(z) \bigr]^*, \label{AnContC} \\[4pt]
 y_l(-z^*) = & \; (-1)^{l+1} \bigl[ y_l(z) \bigr]^*, \label{AnContD} \\[4pt]
 h_l^{(\alpha)}(-z^*) = & \; (-1)^l \bigl[  h_l^{(\alpha)}(z) \bigr]^*, \qquad (\alpha =1,2), \label{AnContE}\\[4pt]
\bigl[ h^{(1)}_l(z) \bigr]^* = & \; h^{(2)}_l(z^*). \label{AnContF}
\end{align}
\end{subequations}
 \item Integrals:
\begin{subequations}\label{Int}
\begin{align}
\int_0^a & j_l(x  \, r) j_{l}(y \, r)\, r^2 \di r  \nonumber \\
 & =  \frac{a^2 }{x^2-y^2}\Bigl[ y \, j_{l}(x a) j_{l-1}(y a)- x \,j_{l-1}(x a) j_{l}(y a)\Bigr],  \label{Int1}
\end{align}
\begin{multline}
\int_0^a j_l^2(x  \, r)  \, r^2 \di r \\ = \frac{a^3 }{2}\Bigl[j_{l}^{\,2}(x a) - j_{l-1}(x a) j_{l+1}(x a) \Bigl]. \label{Int2}
\end{multline}
\end{subequations}
\end{enumerate}
Relations 1. and 2. also hold for $y_l(z)$, $h_l^{(1)}(z)$ and $h_l^{(2)}(z)$.

\section{TE and TM resonances of a dielectric sphere}\label{A}

\begin{figure}[!ht]
\centerline{\includegraphics[scale=3,clip=false,width=1\columnwidth,trim = 0 0 0 0]{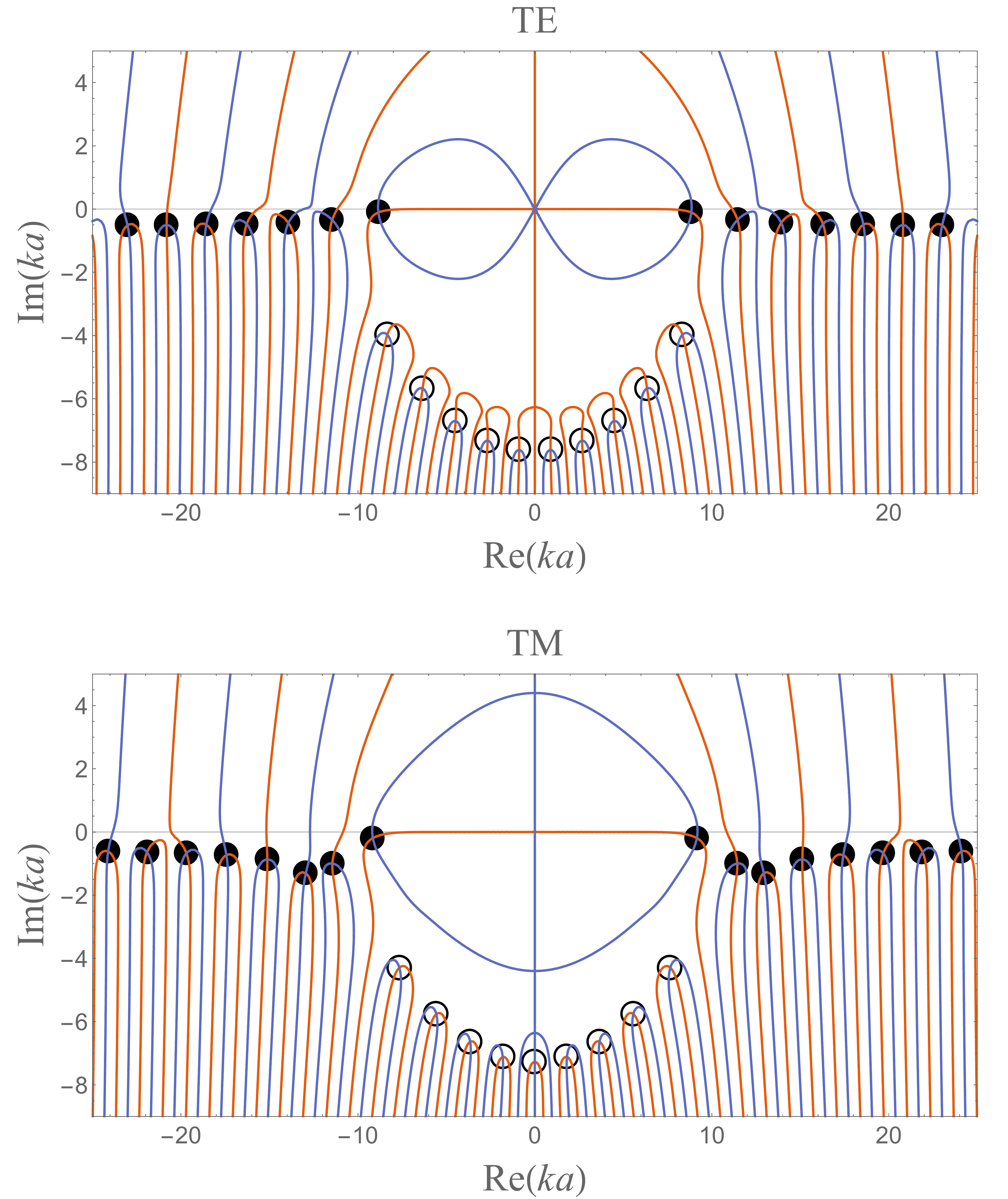}}
\caption{\label{fig1}   Contour plots in the complex $k$-plane of the zeros of $f_l(z)$ for TE (top) and TM (bottom) polarization. Points on the red curves are solutions of the equation $\re f_l(z) =0$ and points on the blue curves are solutions of $\im f_l(z)=0$. The characteristic values $z_{nl}$ are those points where  red and  blue lines cross each other. Filled circles  mark resonant values $z^\text{r}$, open circles indicate non-resonant values $z^\text{nr}$. In this figure, $n_1 = 1.5$ and $l=10$. Values of $k$ with larger real part correspond to higher radial quantum number, where one expects the modes to become more lossy. }
\end{figure}

\begin{figure}[!ht]
\centerline{\includegraphics[scale=3,clip=false,width=1\columnwidth,trim = 0 0 0 0]{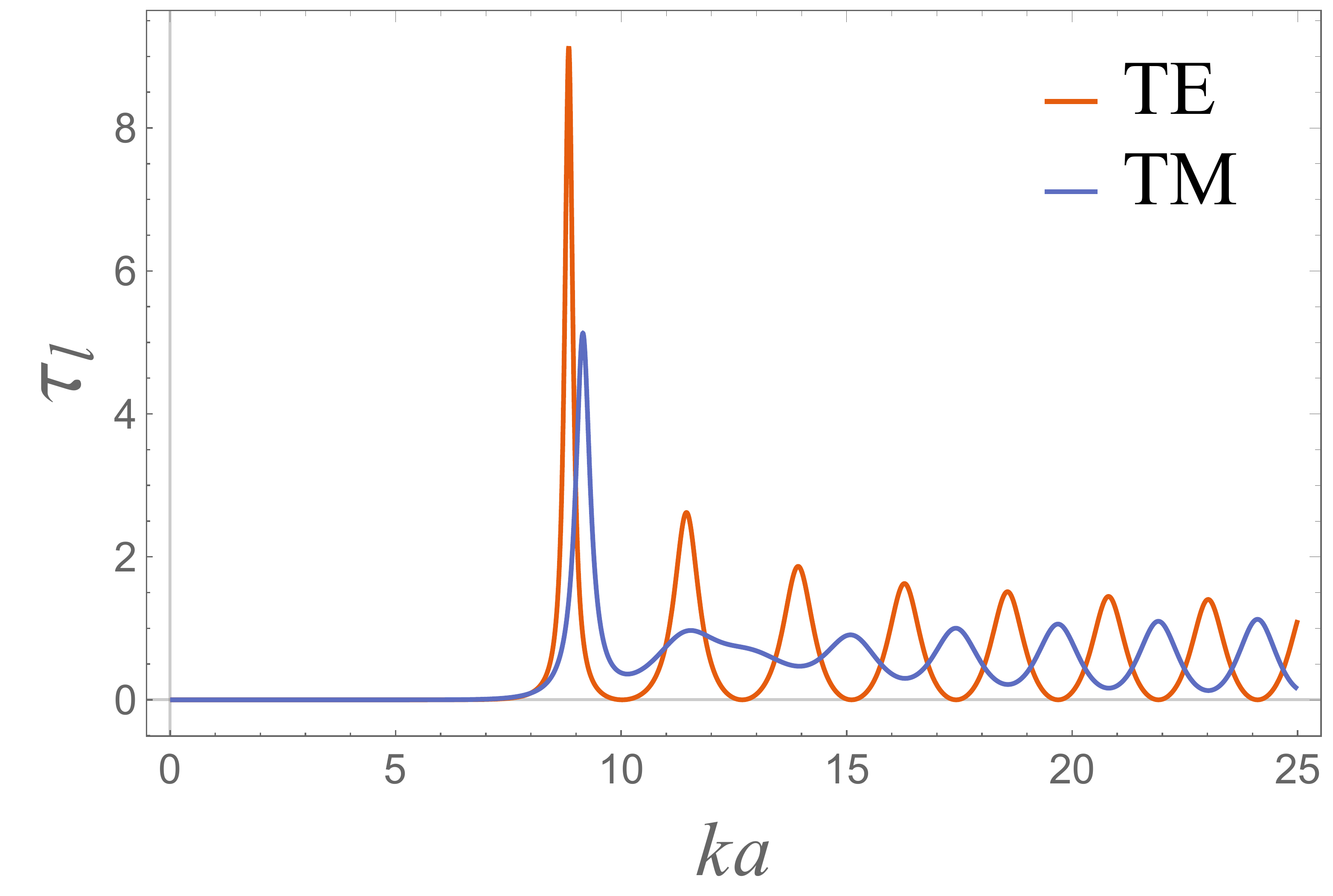}}
\caption{\label{fig2}   Plot of the specific time delay \eqref{TimeDelay} for a glass sphere of radius $a$, refractive index $n_1=1.5$ and azimuthal number $l=10$. Red curve: TE polarization, blue curve: TM  polarization. The peaks of these curves are located at $ka = \re z^\text{r}$, where $z^\text{r}$ denotes a solution of the equation \eqref{Poles} associated with a resonant mode of the field. }
\end{figure}

\begin{figure}[!hb]
\centerline{\includegraphics[scale=3,clip=false,width=1\columnwidth,trim = 0 0 0 0]{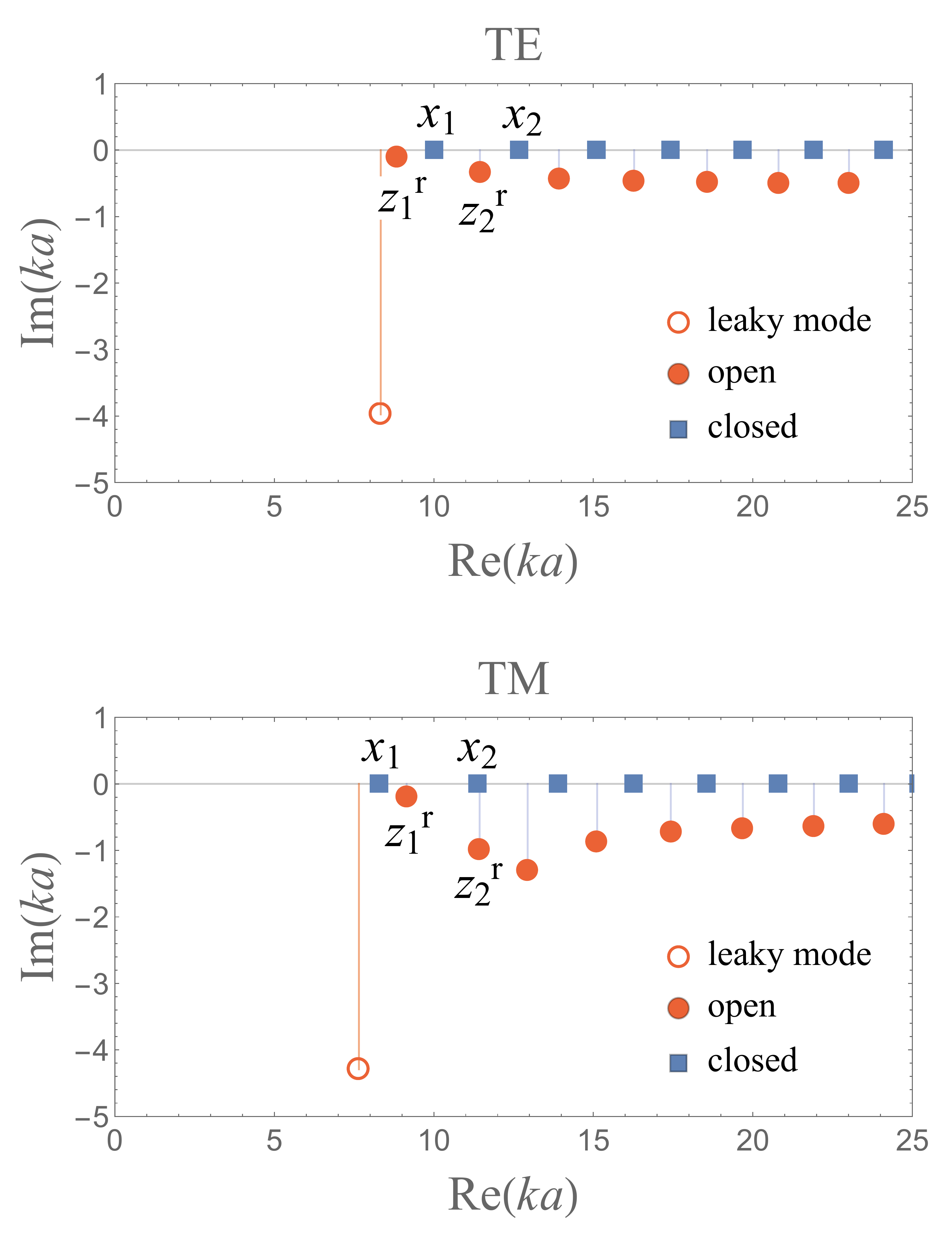}}
\caption{\label{fig3}   Characteristic values for eqs. \eqref{PolesApp} (filled and open red circles) and \eqref{PolesC} (blue squares on the real axis) as given in Fig. \ref{fig1} for $\re (ka) >0$. Top: TE polarization; bottom: TM polarization. The first two resonances  of an open ($z_{1l}^\mathrm{r}$ and $z_{2l}^\mathrm{r}$) and a closed ($x_{1l}$ and $x_{2l}$) sphere  are marked.
In this figure we have chosen the same parameters as in Fig. \ref{fig1}, that is $n_1 = 1.5$ and $l=10$. For the sake of clarity we have omitted the index $l$ in the plots.}
\end{figure}

The complex-valued resonances of the sphere are generally associated with the poles of the scattering amplitudes \eqref{ScatMat}. These poles are found by solving with respect to the complex variable $z = a \left( k' + i k'' \right) \equiv x + i y$ the transcendental equation
\begin{align}\label{PolesApp}
f_l(z) = & \; p \, j_l(n_1 z) \bigl[ z h^{(1)}_l(z) \bigr]' - h^{(1)}_l(z)\bigl[(n_1 z) j_l(n_1 z) \bigr]' \nonumber \\[6pt]
 =  & \;  0,
\end{align}
where $p =1$ for TE polarization and $p = n_1^2/n_2^2$ for  TM polarization. Equation \eqref{Poles} admits solutions only for certain \emph{characteristic values} of the complex variable $z$. These characteristic values form a denumerable set $\{z_{0l}, z_{\pm 1 l} , z_{\pm 2 l},  \ldots, z_{ \pm n l}, \ldots\}$, where $\re z_{nl} = - \re z_{-n l} >0$ and $\im z_{nl} < 0$, for all $n$. We label the poles with $\re k <0$ by the negative index $-n$, so that $k_{-nl} = - k_{nl}^*$. For $l$ odd (even) and TE (TM) polarization there exists a  pole denoted $k_{0l}$ such that $\re k_{0l} = 0$ and $\im k_{0l} < 0$.
Figure \ref{fig1} shows two typical distributions, symmetric with respect to the vertical axis, for TE and TM polarization,  of the roots of $f_l(z)$ in the complex $k$-plane  for a glass sphere of radius $a$, refractive index $n_1=1.5$ and \emph{azimuthal number} $l=10$.
%
%
Filled and open black circles mark, respectively, characteristic values $z^\text{r}$ and $z^\text{nr}$ associated  with resonant and not resonant modes of the field. The latter are very leaky modes that are sometimes called \emph{external whispering gallery modes} \cite{PhysRevA.77.013804}. Grandy \cite{Grandy} suggested that to distinguish between resonant and not resonant characteristic values one should evaluate,
\begin{enumerate*}[label=(\roman*)]
\item the phase shift $\delta_l$,
\item the scattering strength $\sin^2 \delta_l$,
\item the interior wave amplitude $\abs{A_l}$, and
\item the \emph{specific time delay}
\end{enumerate*}
\begin{align}\label{TimeDelay}
\tau_l \equiv \frac{1}{a} \frac{\di\delta_l}{\di k} = \frac{1}{2 \, i} \frac{\di}{\di (k a)}\log S_l.
\end{align}
However, it is possible to show that these conditions are almost equivalent \cite{Ohanian74} and that, for example, it is sufficient to verify the presence of a sharp peak in  $\tau_l$ per each value of $ka = \re z^\text{r}$, as shown in Fig. \ref{fig2}.
%
%
In practice,  we wish   $z_{1l}^\textrm{r}$ to be the pole with $\re z_{1l}^\textrm{r} \sim l/n_1$ and the smallest imaginary part (the subsequent resonant values will be ordered according to  $\re z_{1l}^\textrm{r}<\re z_{2l}^\textrm{r}< \ldots$, et cetera.)  Therefore,
the resonant  $z_{nl}^\textrm{r}$  can be found by comparing the solutions of \eqref{Poles} with the characteristic values of TE and TM modes of the \textbf{same} dielectric sphere but embedded in a medium of infinite conductivity (closed sphere), having these values null imaginary parts. They  are the real-valued solutions $\{ x_{1l} , x_{2l},  \ldots, x_{nl}, \ldots\}$, with $x_{1l}< x_{2l} < \ldots$, of  \cite{Stratton}:
\begin{align}\label{PolesC}
\left\{  \begin{array}{ll}
                       j_l(n_1 x)=0, & \quad \text{TE polarization,} \\[6pt]
                       \bigl[(n_1 x) j_l(n_1 x) \bigr]' =0, & \quad \text{TM polarization.}
                     \end{array} \right.
\end{align}
Thus, we \emph{define} $z_{1l}^\textrm{r}$ as  the solution of $f_l(z) = 0$ closest  to the smaller root  $x_{1l}$  of \eqref{PolesC}, namely
\begin{align}\label{thumb}
\! \! \! z_{1l}^\textrm{r} = \{z \in \mathbb{C}  | \, f_l(z) = 0, \, \re z >0, \, \abs{z-x_{1l}} = \min \}.
\end{align}
Then, given a solution $z$ of $f_l(z) = 0$,  $z = z^\text{r}$ if either $\re z > \re z_{1l}^\textrm{r} $ or $\re z <- \re z_{1l}^\textrm{r} $, because $z$ and $-z^*$ are solutions of the same equation.
This empirical rule is illustrated in Fig. \ref{fig3}, which shows the location in the complex $k$-plane a few solutions of \eqref{PolesApp} (filled and open red circles) and \eqref{PolesC} (blue squares on the real axis) for a glass sphere of radius $a$, refractive index $n_1=1.5$ and \emph{azimuthal number} $l=10$.
%
%

\section{Perturbative diagonalization of $H_l$}\label{Hldiag}

As $\Delta_\text{rip} \ll e^2$,  we can diagonalize $H_l$ using perturbation theory after defining
\begin{align}\label{Deg10c}
H_l = H_l^{(0)} + H_l^{(1)},
\end{align}
where
\begin{equation}\label{Deg15b}
\begin{split}
[H_l^{(0)} ]_{m m'} = & \; ({Y}_{lm} , h_\text{rot}(\hbr) {Y}_{lm'})_\hbr = \delta_{m m'} \, \ell^{(0)}_{l\abs{m}},  \\[4pt]
[H_l^{(1)} ]_{m m'}  = & \; ({Y}_{lm} , h_\text{rip}(\hbr) {Y}_{lm'})_\hbr,
\end{split}
\end{equation}
with $\ell^{(0)}_{l\abs{m}}$  defined by \eqref{Deg9}. If $m \neq 0$ each eigenvalue of $H_l^{(0)}$ is doubly degenerate
%
and \eqref{Deg7} can be approximately rewritten as:
\begin{align}\label{Deg10b}
\ell_{lm} \approx  \ell^{(0)}_{l\abs{m}} + \ell^{(1)}_{l,\pm \abs{m}},
\end{align}
where  $ \ell^{(1)}_{l,+\abs{m}}$ and $  \ell^{(1)}_{l,-\abs{m}} $ denote the eigenvalues of the $2 \times 2$ matrix $H_{lm}^{(1)}$ defined by
\begin{multline}\label{Deg11}
H_{l m }^{(1)}
\equiv
\begin{bmatrix}
  \bigl({Y}_{l m } , h_\text{rip}(\hbr){Y}_{l m } \bigr)_\hbr & \bigl({Y}_{l m } , h_\text{rip}(\hbr) {Y}_{l,- m } \bigr)_\hbr \\[6pt]
  \bigl({Y}_{l,- m } , h_\text{rip}(\hbr) {Y}_{l m } \bigr)_\hbr & \bigl({Y}_{l,- m } , h_\text{rip}(\hbr) {Y}_{l,- m } \bigr)_\hbr\\[3pt]
\end{bmatrix},
\end{multline}
with  $m=1,2,\ldots, l$. Explicitly,
\begin{align}\label{Deg12}
\ell^{(1)}_{l,\pm \abs{m}} = \frac{1}{2} \bigl( H_{11} + H_{22} \pm \Delta \bigr),
\end{align}
where here and hereafter we use the shorthand notation $[H_{lm}^{(1)}]_{ij} \equiv H_{ij}, \,(i,j=1,2) $, and
\begin{align}\label{Deg13}
\Delta = \sqrt{( H_{11} - H_{22})^2 + 4 \abs{H_{12}}^2}.
\end{align}
 If $m=0$ we can apply non-degenerate perturbation theory to obtain
\begin{align}\label{Deg11b}
\ell_{l0} \approx \ell^{(0)}_{l0} + \bigl({Y}_{l 0 } , h_\text{rip}(\hbr){Y}_{l 0 } \bigr)_\hbr.
\end{align}
Similarly, the functions $\mathcal{Y}_{lm}(\hbr)$ spanning the degenerate subspace $\mathcal{D}_{nl}$ can be approximated by
\begin{align}\label{Deg11c}
\mathcal{Y}_{lm}(\hbr) \approx Y_{lm}(\hbr) + (1-\delta_{m0}) \mathcal{Y}_{l,\pm \abs{m}}^{(0)}(\hbr),
\end{align}
where
\begin{align}\label{Deg14}
  \mathcal{Y}_{l,\pm \abs{m}}^{(0)}(\hbr) = C_{1\pm}  Y_{l\abs{m}}(\hbr) + C_{2\pm}  Y_{l,-\abs{m}}(\hbr),
\end{align}
with the superscript ``$(0)$'' marking the zero-order character of these corrections, and
\begin{equation}\label{Deg15}
\begin{split}
C_{1\pm} = & \; \sqrt{\frac{H_{12}}{2 \abs{H_{12}}}\left(1 \pm \frac{H_{11} - H_{22}}{\Delta} \right)},  \\[4pt]
C_{2\pm} = & \; \pm \sqrt{\frac{H_{21}}{2 \abs{H_{21}}}\left(1 \mp \frac{H_{11} - H_{22}}{\Delta} \right)} .
\end{split}
\end{equation}
%


\end{document}